\documentclass[12pt,preprint]{aastex62}

\usepackage{graphics,graphicx,natbib}
\usepackage{hyperref}
\def\lapp{\ifmmode\stackrel{<}{_{\sim}}\else$\stackrel{<}{_{\sim}}$\fi}
\def\gapp{\ifmmode\stackrel{>}{_{\sim}}\else$\stackrel{>}{_{\sim}}$\fi}

\newcommand{\fix}[1]{{\textcolor{black} { #1 }}}
\begin{document}

\title{THE CHIME FAST RADIO BURST PROJECT:  SYSTEM OVERVIEW}

\author{The CHIME/FRB Collaboration:  M.~Amiri}
  \affiliation{Department of Physics and Astronomy, University of British Columbia, 6224 Agricultural Road, Vancouver, BC V6T 1Z1, Canada}
\author{K.~Bandura}
  \affiliation{Center for Gravitational Waves and Cosmology, West Virginia University, Chestnut Ridge Research Building, Morgantown, WV 26505, USA}
  \affiliation{CSEE, West Virginia University, Morgantown, WV 26505, USA}
\author{P.~Berger }
  \affiliation{Canadian Institute for Theoretical Astrophysics, University of Toronto, 60 St.~George Street, Toronto, ON M5S 3H8, Canada}
  \affiliation{Department of Physics, University of Toronto, 60 St.~George St, Toronto, ON M5S 3H4, Canada}
\author{M.~Bhardwaj}
  \affiliation{Department of Physics, McGill University, 3600 rue University, Montr\'eal, QC H3A 2T8, Canada}
  \affiliation{McGill Space Institute, McGill University, 3550 rue University, Montr\'eal, QC H3A 2A7, Canada}
\author{M.~M.~Boyce}
  \affiliation{Department of Physics, McGill University, 3600 rue University, Montr\'eal, QC H3A 2T8, Canada}
  \affiliation{McGill Space Institute, McGill University, 3550 rue University, Montr\'eal, QC H3A 2A7, Canada}
\author{P.~J.~Boyle}
  \affiliation{Department of Physics, McGill University, 3600 rue University, Montr\'eal, QC H3A 2T8, Canada}
  \affiliation{McGill Space Institute, McGill University, 3550 rue University, Montr\'eal, QC H3A 2A7, Canada}
\author{C.~Brar}
  \affiliation{Department of Physics, McGill University, 3600 rue University, Montr\'eal, QC H3A 2T8, Canada}
  \affiliation{McGill Space Institute, McGill University, 3550 rue University, Montr\'eal, QC H3A 2A7, Canada}
\author{M.~Burhanpurkar}
  \affiliation{Harvard University, Cambridge MA, 02138, USA}
\author{P.~Chawla}
  \affiliation{Department of Physics, McGill University, 3600 rue University, Montr\'eal, QC H3A 2T8, Canada}
  \affiliation{McGill Space Institute, McGill University, 3550 rue University, Montr\'eal, QC H3A 2A7, Canada}
\author{J.~Chowdhury}
  \affiliation{Department of Physics, McGill University, 3600 rue University, Montr\'eal, QC H3A 2T8, Canada}
  \affiliation{McGill Space Institute, McGill University, 3550 rue University, Montr\'eal, QC H3A 2A7, Canada}
  \affiliation{Ming Hsieh Department of Electrical Engineering, University of Southern California, 3740 McClintock Avenue, EEB 102, Los Angeles, CA 90089-2560, USA}
\author{J.-F.~Cliche}
  \affiliation{Department of Physics, McGill University, 3600 rue University, Montr\'eal, QC H3A 2T8, Canada}
  \affiliation{McGill Space Institute, McGill University, 3550 rue University, Montr\'eal, QC H3A 2A7, Canada}
\author{M.~D.~Cranmer}
  \affiliation{Department of Physics, McGill University, 3600 rue University, Montr\'eal, QC H3A 2T8, Canada}
  \affiliation{McGill Space Institute, McGill University, 3550 rue University, Montr\'eal, QC H3A 2A7, Canada}
\author{D.~Cubranic}
  \affiliation{Department of Physics and Astronomy, University of British Columbia, 6224 Agricultural Road, Vancouver, BC V6T 1Z1, Canada}
\author{M.~Deng}
  \affiliation{Department of Physics and Astronomy, University of British Columbia, 6224 Agricultural Road, Vancouver, BC V6T 1Z1, Canada}
\author{N.~Denman}
  \affiliation{Department of Astronomy and Astrophysics, University of Toronto, 50 St.~George Street, Toronto, ON M5S 3H4, Canada}
  \affiliation{Dunlap Institute for Astronomy and Astrophysics, University of Toronto, 50 St.~George Street, Toronto, ON M5S 3H4, Canada}
\author{M.~Dobbs}
  \affiliation{Department of Physics, McGill University, 3600 rue University, Montr\'eal, QC H3A 2T8, Canada}
  \affiliation{McGill Space Institute, McGill University, 3550 rue University, Montr\'eal, QC H3A 2A7, Canada}
\author{M.~Fandino}
  \affiliation{Department of Physics and Astronomy, University of British Columbia, 6224 Agricultural Road, Vancouver, BC V6T 1Z1, Canada}
\author{E.~Fonseca}
  \affiliation{Department of Physics, McGill University, 3600 rue University, Montr\'eal, QC H3A 2T8, Canada}
  \affiliation{McGill Space Institute, McGill University, 3550 rue University, Montr\'eal, QC H3A 2A7, Canada}
\author{ B.~M.~Gaensler}
  \affiliation{Dunlap Institute for Astronomy and Astrophysics, University of Toronto, 50 St.~George Street, Toronto, ON M5S 3H4, Canada}
\author{U.~Giri}
  \affiliation{Perimeter Institute for Theoretical Physics, 31 Caroline Street N, Waterloo, ON N2L 2Y5, Canada}
\author{A.~J.~Gilbert}
  \affiliation{Department of Physics, McGill University, 3600 rue University, Montr\'eal, QC H3A 2T8, Canada}
  \affiliation{McGill Space Institute, McGill University, 3550 rue University, Montr\'eal, QC H3A 2A7, Canada}
\author{D.~C.~Good}
  \affiliation{Department of Physics and Astronomy, University of British Columbia, 6224 Agricultural Road, Vancouver, BC V6T 1Z1, Canada}
\author{S.~Guliani}
  \affiliation{Department of Physics and Astronomy, University of British Columbia, 6224 Agricultural Road, Vancouver, BC V6T 1Z1, Canada}
\author{M.~Halpern}
  \affiliation{Department of Physics and Astronomy, University of British Columbia, 6224 Agricultural Road, Vancouver, BC V6T 1Z1, Canada}
\author{G.~Hinshaw}
  \affiliation{Department of Physics and Astronomy, University of British Columbia, 6224 Agricultural Road, Vancouver, BC V6T 1Z1, Canada}
\author{C.~H\"ofer }
  \affiliation{Department of Physics and Astronomy, University of British Columbia, 6224 Agricultural Road, Vancouver, BC V6T 1Z1, Canada}
\author{A.~Josephy}
  \affiliation{Department of Physics, McGill University, 3600 rue University, Montr\'eal, QC H3A 2T8, Canada}
  \affiliation{McGill Space Institute, McGill University, 3550 rue University, Montr\'eal, QC H3A 2A7, Canada}
\author{V.~M.~Kaspi}
  \affiliation{Department of Physics, McGill University, 3600 rue University, Montr\'eal, QC H3A 2T8, Canada}
  \affiliation{McGill Space Institute, McGill University, 3550 rue University, Montr\'eal, QC H3A 2A7, Canada}
\author{T.~L.~Landecker}
  \affiliation{Dominion Radio Astrophysical Observatory, Herzberg Astronomy \& Astrophysics Research Centre, National Reseach Council of Canada, P.O. Box 248, Penticton, V2A 6J9, Canada}
\author{D.~Lang}
  \affiliation{Department of Astronomy and Astrophysics, University of Toronto, 50 St.~George Street, Toronto, ON M5S 3H4, Canada}
  \affiliation{Dunlap Institute for Astronomy and Astrophysics, University of Toronto, 50 St.~George Street, Toronto, ON M5S 3H4, Canada}
  \affiliation{Perimeter Institute for Theoretical Physics, 31 Caroline Street N, Waterloo, ON N2L 2Y5, Canada}
\author{H.~Liao}
  \affiliation{Department of Physics, McGill University, 3600 rue University, Montr\'eal, QC H3A 2T8, Canada}
\author{K.~W.~Masui}
  \affiliation{Department of Physics and Astronomy, University of British Columbia, 6224 Agricultural Road, Vancouver, BC V6T 1Z1, Canada}
\author{J.~Mena-Parra}
  \affiliation{Department of Physics, McGill University, 3600 rue University, Montr\'eal, QC H3A 2T8, Canada}
  \affiliation{McGill Space Institute, McGill University, 3550 rue University, Montr\'eal, QC H3A 2A7, Canada}
\author{A.~Naidu}
  \affiliation{Department of Physics, McGill University, 3600 rue University, Montr\'eal, QC H3A 2T8, Canada}
  \affiliation{McGill Space Institute, McGill University, 3550 rue University, Montr\'eal, QC H3A 2A7, Canada}
\author{L.~B.~Newburgh}
  \affiliation{Department of Physics, Yale University, New Haven, CT 06520, USA}
\author{C.~Ng}
  \affiliation{Dunlap Institute for Astronomy and Astrophysics, University of Toronto, 50 St.~George Street, Toronto, ON M5S 3H4, Canada}
\author{C.~Patel}
  \affiliation{Department of Physics, McGill University, 3600 rue University, Montr\'eal, QC H3A 2T8, Canada}
\author{U.-L.~Pen}
  \affiliation{Canadian Institute for Theoretical Astrophysics, University of Toronto, 60 St.~George Street, Toronto, ON M5S 3H8, Canada}
  \affiliation{Dunlap Institute for Astronomy and Astrophysics, University of Toronto, 50 St.~George Street, Toronto, ON M5S 3H4, Canada}
  \affiliation{Perimeter Institute for Theoretical Physics, 31 Caroline Street N, Waterloo, ON N2L 2Y5, Canada}
\author{T.~Pinsonneault-Marotte}
  \affiliation{Department of Physics and Astronomy, University of British Columbia, 6224 Agricultural Road, Vancouver, BC V6T 1Z1, Canada}
\author{Z.~Pleunis}
  \affiliation{Department of Physics, McGill University, 3600 rue University, Montr\'eal, QC H3A 2T8, Canada}
  \affiliation{McGill Space Institute, McGill University, 3550 rue University, Montr\'eal, QC H3A 2A7, Canada}
\author{M.~Rafiei Ravandi}
  \affiliation{Perimeter Institute for Theoretical Physics, 31 Caroline Street N, Waterloo, ON N2L 2Y5, Canada}
\author{S.~M.~Ransom}
  \affiliation{NRAO, 520 Edgemont Rd., Charlottesville, VA 22903, USA}
\author{A.~Renard}
  \affiliation{Dunlap Institute for Astronomy and Astrophysics, University of Toronto, 50 St.~George Street, Toronto, ON M5S 3H4, Canada}
\author{P.~Scholz}
  \affiliation{Dominion Radio Astrophysical Observatory, Herzberg Astronomy \& Astrophysics Research Centre, National Reseach Council of Canada, P.O. Box 248, Penticton, V2A 6J9, Canada}
  \author{K. Sigurdson}
  \affiliation{Department of Physics and Astronomy, University of British Columbia, 6224 Agricultural Road, Vancouver, BC V6T 1Z1, Canada}
\author{S.~R.~Siegel}
  \affiliation{Department of Physics, McGill University, 3600 rue University, Montr\'eal, QC H3A 2T8, Canada}
  \affiliation{McGill Space Institute, McGill University, 3550 rue University, Montr\'eal, QC H3A 2A7, Canada}
\author{K.~M.~Smith}
  \affiliation{Perimeter Institute for Theoretical Physics, 31 Caroline Street N, Waterloo, ON N2L 2Y5, Canada}
\author{I.~H.~Stairs}
  \affiliation{Department of Physics and Astronomy, University of British Columbia, 6224 Agricultural Road, Vancouver, BC V6T 1Z1, Canada}
\author{S.~P.~Tendulkar}
  \affiliation{Department of Physics, McGill University, 3600 rue University, Montr\'eal, QC H3A 2T8, Canada}
  \affiliation{McGill Space Institute, McGill University, 3550 rue University, Montr\'eal, QC H3A 2A7, Canada}
\author{K.~Vanderlinde}
  \affiliation{Department of Astronomy and Astrophysics, University of Toronto, 50 St.~George Street, Toronto, ON M5S 3H4, Canada}
  \affiliation{Dunlap Institute for Astronomy and Astrophysics, University of Toronto, 50 St.~George Street, Toronto, ON M5S 3H4, Canada}
\author{D.~V.~Wiebe}
  \affiliation{Department of Physics and Astronomy, University of British Columbia, 6224 Agricultural Road, Vancouver, BC V6T 1Z1, Canada}
  
  \correspondingauthor{V.~M.~Kaspi}
  \email{vkaspi@physics.mcgill.ca}

\begin{abstract}
The Canadian Hydrogen Intensity Mapping Experiment (CHIME) is a novel transit radio
telescope operating across the 400--800-MHz band.  CHIME is comprised of four 20-m $\times$ 100-m semi-cylindrical paraboloid reflectors, each of which has 256 dual-polarization feeds suspended along its axis, giving it a $\gapp$ 200 square degree field-of-view. This, combined with wide bandwidth, high sensitivity, and a powerful correlator makes CHIME an excellent instrument for the detection of Fast Radio Bursts (FRBs). The CHIME Fast Radio Burst Project (CHIME/FRB) will search beam-formed, high time- and frequency-resolution data in real time for FRBs in the CHIME field-of-view. Here we describe the CHIME/FRB backend, including the real-time FRB search and detection software pipeline as well as the planned offline analyses. 
\fix{We estimate a CHIME/FRB detection rate of 2--42 FRBs/sky/day normalizing to the rate estimated at 1.4-GHz by \citet{vbl+16}.  Likely science outcomes of CHIME/FRB are also discussed.}
CHIME/FRB is currently operational in a commissioning phase, with science operations expected
to commence in the latter half of 2018.
\end{abstract}


   \section{Fast Radio Bursts}
\label{sec:intro}
The Fast Radio Burst (FRB) phenomenon was first recognized when \citet{lbm+07} discovered a bright (30 Jy) radio burst of duration $<5$~ms
well off the Southern Galactic plane,
in a search for single pulses using the Parkes radio
telescope.
The burst had a dispersion measure (DM) of
375~pc~cm$^{-3}$, far in excess of the maximum expected along this
line of sight from the Galaxy (25~pc~cm$^{-3}$) according to the
 NE2001 model for the Galactic electron density distribution \citep{cl01}.
The large excess DM suggests an extragalactic origin, as first noted by \citet{lbm+07}.
Subsequently, \citet{tsb+13} reported four additional FRBs,
also detected using the Parkes telescope, each with a DM greatly
in excess of the total line-of-sight Galactic electron column density. Following this, FRBs became widely accepted as a newly recognized astrophysical phenomenon.
The identification of particularly unusual FRB-like ``peryton"
events with radio interference local to the Parkes
telescope \citep{pkb+15} ultimately solidified the astrophysical nature
of the remaining, non-peryton sources.  The discovery of an FRB using
the Arecibo Observatory\footnote{\url{www.naic.edu}} \citep{sch+14} laid to rest concerns regarding FRBs being a telescope-specific phenomenon.  Since then, one FRB has
been discovered using the Green Bank Telescope \citep{mls+15a}, and an
additional two dozen have been found using Parkes, the UTMOST telescope
\citep[e.g.][]{cfb+17} as well as ASKAP \citep{bsm+17}.  This brings the
total number of formally reported FRBs\footnote{See the online FRB
catalog at \url{www.frbcat.org}.} to approximately 30.
Of these, most were discovered at a radio frequency of 1.4 GHz, except
the GBT and UTMOST events which were detected near 800 MHz.

Although the number of detected FRBs today is fewer than three dozen, the inferred sky rate at 1.4 GHz is estimated to be between several hundred and a few thousand per sky per day \citep[e.g.][]{tsb+13,rlb+16,lvl+17} at existing fluence thresholds ($\sim 1\,\mathrm{Jy\,ms}$). The discrepancy between the high inferred rate and the relatively small number of detections
is largely due to the small fields-of-view (FoVs) of existing telescopes.  There are indications that the overall rate is substantially lower at frequencies below $\sim$400 MHz \citep{kca+15,rbm+16,ckj+17}.  

Assuming standard estimates for the intergalactic free electron density \citep[e.g.][]{ino04}, as well as reasonable estimates for the DM contribution from host galaxies, the events appear to be from cosmological distances (redshifts $z\sim0.1-2$).  The estimated volumetric FRB rate does not match with those measured for other known transient events such as gamma-ray bursts or supernovae \citep{kon+14}. The sources of FRBs are thus currently unknown, but models typically include compact objects, due to the short timescales involved, with some invoking cataclysmic events \citep[see][for a recent review]{rl17}.  Being thus far discovered primarily using diffraction-limited single-dish radio telescopes at decimetre wavelengths, or interferometers with limited baselines, FRB positional uncertainties are typically many arcminutes, prohibiting identifications of multi-wavelength counterparts or host galaxies. 

Thus far, only the Arecibo-discovered FRB 121102 has been observed to repeat \citep{ssh+16a,ssh+16b}. This discovery demonstrates that at least this particular source does not have its origin in a cataclysmic event. Many dozen repeat bursts, all at a consistent DM, have now been detected and show strong evidence for clustering in time \citep{ssh+16b,clw+17,op17,msh+18}. Detection of multiple bursts interferometrically enable a subarcsecond localization \citep{clw+17,mph+17} in a dwarf galaxy at redshift $z\approx0.2$ \citep{tbc+17}. This provides strong support for 
the interpretation that the large excess DMs of other FRBs imply cosmological distances and suggests that FRBs may be useful new probes of galaxy halos, large-scale structure, the intergalactic medium, and of cosmological parameters \citep[e.g.][]{mcq14,ms15,rsb+16,yz16,fl16,pn18,yt18,sd18}.

Whether all FRBs repeat is currently unknown.  Apart from FRB 121102, observations of several FRB positions for many hours \citep[e.g.][]{pjk+15} have not revealed repeated
bursts from the same source.  In one case, a second event, FRB 140514, was observed from the same arcminute-area beam on the sky as FRB 110220,
but at a very different DM. Petroff et al. conclude they are likely unrelated \citep[but see][for an alternative discussion]{mls+15b}. Thus, whether the repeating nature of FRB 121102 is unique is unknown, as are many basic characteristics describing the population, such as their true sky distribution, the luminosity, DM, or scattering time distributions, and whether any of these properties are correlated.  

\begin{figure}[t]
	chimephoto\center{\includegraphics[width=\textwidth]{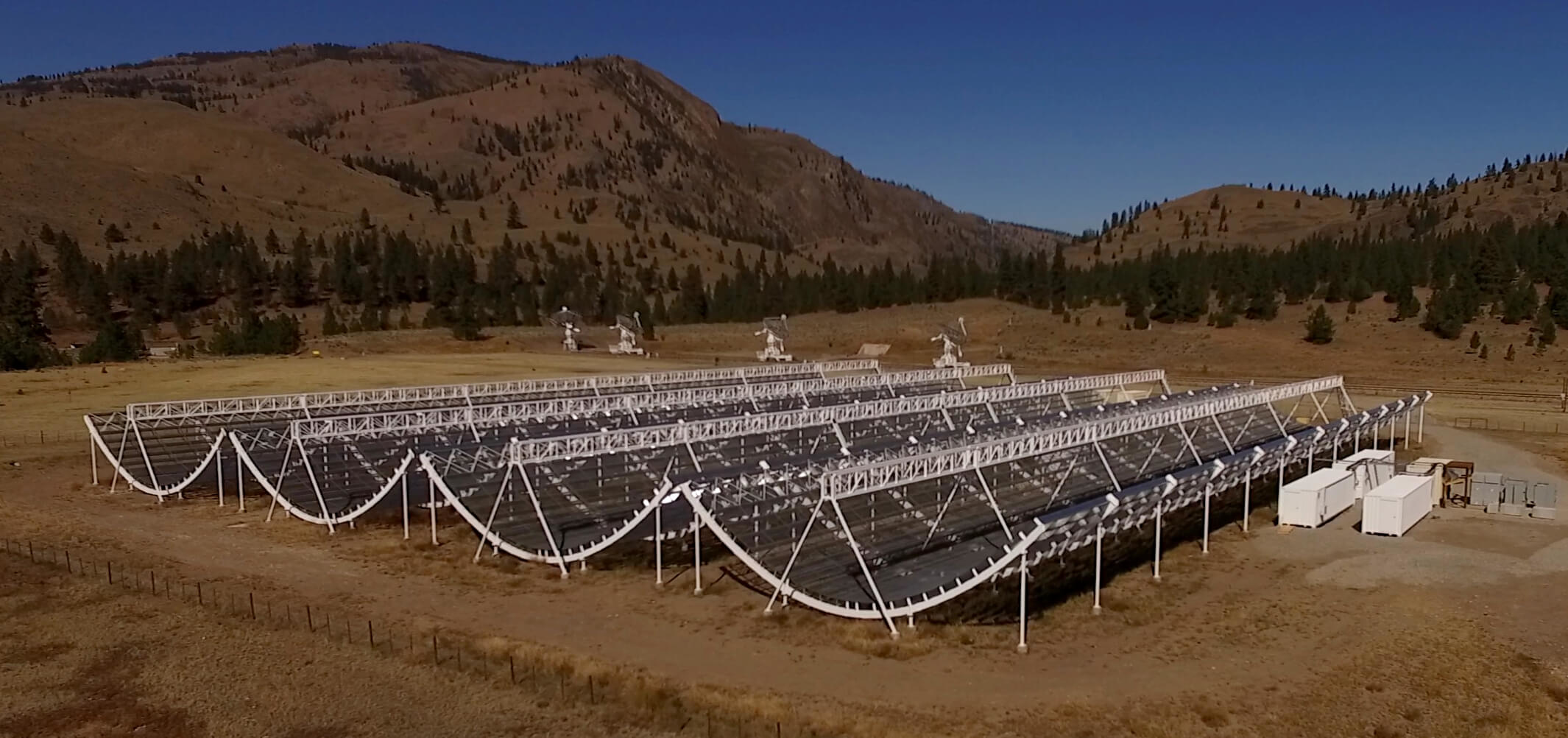}
    \caption{Photograph of the CHIME telescope on 15 September, 2016, looking North-West. The shipping containers housing the X-Engine and CHIME/FRB backend can be seen adjacent to the right-most cylinder. The receiver huts containing the F-Engine are beneath the reflectors and cannot be seen here. The DRAO Synthesis Telescope \citep{klg10} can be seen in the background.  \fix{See Table~\ref{ta:chime} for detailed properties of CHIME.}}
    \label{fig:photo}}
\end{figure}

Here we describe a project underway to enable the Canadian Hydrogen Intensity Mapping Experiment (CHIME)\footnote{\url{www.chime-experiment.ca}} --- originally designed to map baryon acoustic oscillation (BAO) features in redshifted neutral hydrogen gas as a measure of dark energy \citep{naa+14} --- to be, in parallel, a sensitive FRB detector: the CHIME Fast Radio Burst Project (hereafter CHIME/FRB). \fix{CHIME's large $\gapp 200$~sq.~deg. FoV, large collecting area, wide radio bandwidth and powerful correlator, which provides 1024 independent beams within the telescope's primary beam,} will enable this FRB backend to detect multiple FRBs per day, given current estimates of the FRB rate.  The instrument should also detect many Galactic Rotating
Radio Transients (RRATs) \citep{mll+06}. 
Moreover, its sensitivity across the wide 400--800-MHz band is ideal for studying the frequency dependence of the FRB sky rate.  In \S\ref{sec:telescope} we provide a brief description of the CHIME telescope structure and analog system.  \S\ref{sec:correlator} describes the powerful CHIME correlator, whose original design for cosmology was upgraded significantly to accommodate intensity beam-forming for the FRB searches. \S\ref{sec:chimefrb} describes the CHIME/FRB real-time search, detection and analysis procedure. 
Planned analysis features are described in \S\ref{sec:planned}, and the commissioning status of CHIME/FRB is described in \S\ref{sec:commissioning}.  \S\ref{sec:rate} provides an up-to-date estimate of the expected CHIME/FRB event rate, together with a discussion of the science to be probed by CHIME/FRB data.

In addition to the cosmology experiment and CHIME/FRB, the CHIME telescope will also perform daily timing observations of known radio pulsars and RRATs using a dedicated, independent back-end: 
``CHIME/Pulsar."  Information about this latter effort is described by \citet{n++17}.

\section{The CHIME Telescope Structure, Feeds, and Analog Signal Path}
\label{sec:telescope}

\begin{table}[t]
\begin{center}
\caption{Key Properties of the CHIME Telescope Relevant to the CHIME/FRB Project$^a$}
\begin{tabular}{lc} \hline
    Parameter  &  Value \\\hline
    Collecting area & 8000 m$^2$ \\
    Longitude & $119^{\circ}37' 25''.25 $ West \\
    Latitude & $49^{\circ}19' 14''.52$ North\\
    Frequency range & 400--800 MHz \\
    Polarization & orthogonal linear \\
    E-W FoV & 2.5$^{\circ}$--1.3$^{\circ}$ \\
    N-S FoV & $\sim$110$^{\circ}$ \\
    Focal ratio, $f/D$ & 0.25 \\ 
    Receiver noise temperature & 50 K \\
    Number of beams & 1024 \\
    Beam width (FWHM) & 40$^\prime$--20$^\prime$ \\
    FRB search time resolution & 0.983 ms \\
    FRB search frequency resolution & 24.4 kHz \\
    Source transit duration & Equator: 10-–5 min \\
                            & 45$^{\circ}$: 14-–7 min \\
                            & North Celestial Pole:  24 hr \\\hline
\end{tabular}
\label{ta:chime}
\end{center}
$^a$Where two numbers appear, they refer to the low and high frequency edges of the band, respectively.
\end{table}

The CHIME telescope is located on the grounds of the Dominion Radio Astrophysical Observatory (DRAO) near Penticton, British Columbia. The choice of operating frequency, collecting area, and angular resolution for the CHIME telescope was driven by the original motivation for the project: hydrogen intensity mapping of the entire Northern hemisphere to probe the accelerating expansion of the universe over the redshift range where dark energy began to exert its influence, $z = 0.8-2.5$.
Since the BAO signal is weak, and large sky coverage is needed to overcome \fix{sample variance,} exceptionally fast mapping speed is required, driving the instrument to a design with many hundreds of feeds to achieve the mapping goal in a reasonable amount of time.  As 100-m class telescopes are expensive, and no positions on the sky are favored, a transit telescope with no moving parts is the preferred option. Table~\ref{ta:chime} provides a summary of the telescope's key properties.  Detailed telescope performance metrics will be provided in a future publication.

The CHIME telescope (see Fig.~\ref{fig:photo}) consists of four 20-m wide and 100-m long cylindrical paraboloidal reflectors. The inter-cylinder gaps are 2~m. The cylinders are stationary, aligned North-South. The focal length was chosen to be 5 m (${f/D}=0.25$) to place the feeds in the plane of the aperture to minimize cross-talk and ground pick-up.  With no moving parts, the telescope structure could be built at low cost. The structure is steel, built of conventional sections and components, to the dimensional accuracy common for commercial buildings. The reflecting surface is galvanized steel mesh with 16-mm openings, a compromise between allowing snow to fall through and minimizing ground noise leaking through to the feeds. The measured surface roughness is $\sim$9\,mm ($<$2\% of the observing wavelength), adequate for operation in the CHIME band.

\begin{figure}[t]
	\centering\includegraphics[width=0.9\textwidth]{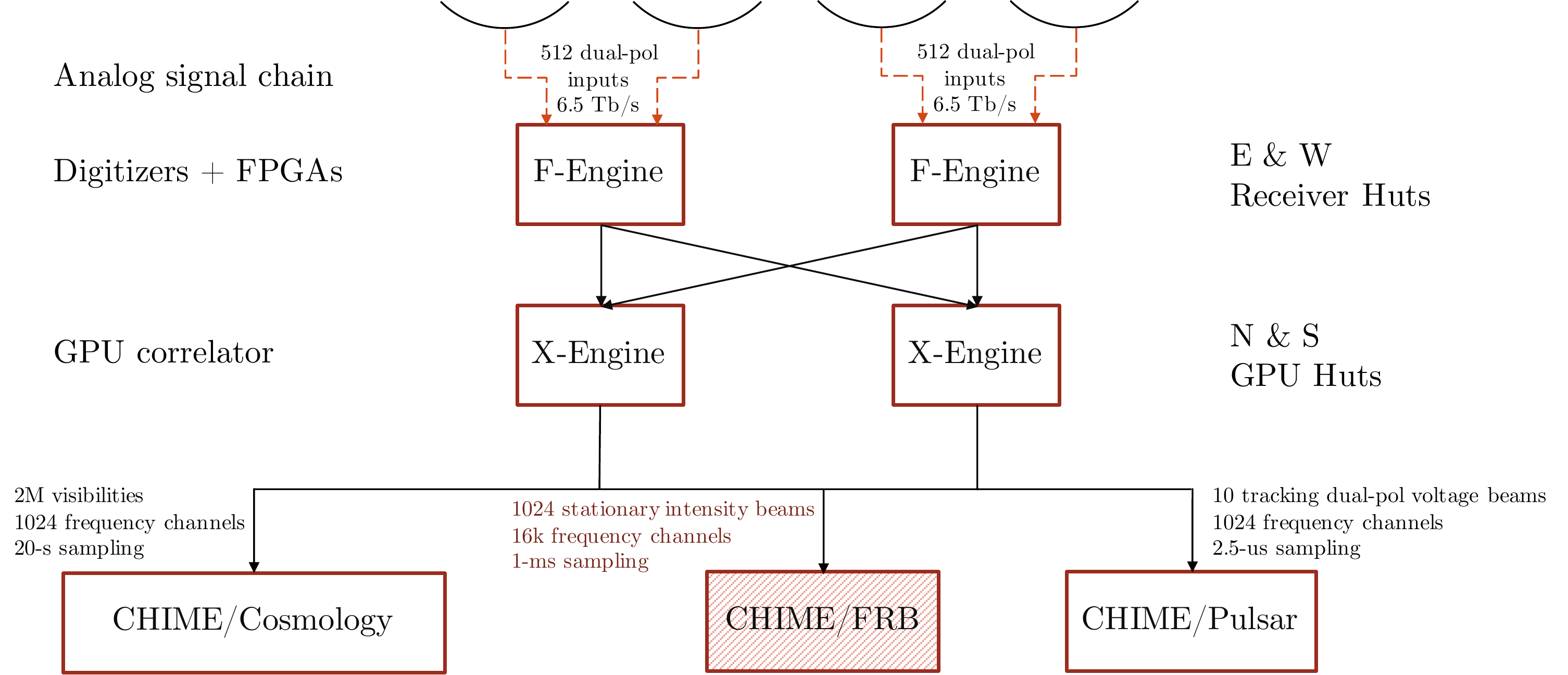}
    \caption{Schematic of the CHIME telescope signal path. The four cylinders (black arcs), the correlator (F- and X-Engines), and the backend science instruments are shown. The dashed orange segments depict analog-signal carrying coaxial cables from the 256 feeds on each cylinder to the F-Engines in the corresponding East or West receiver huts beneath the cylinders. The black segments depict digital data carried through copper and fiber cables. Networking devices are not shown. The X-Engine is housed in two shipping containers (labeled North and South) adjacent to the cylinders. The HI intensity map making (CHIME/Cosmology) and CHIME/Pulsar backends are housed in a shielded room in the DRAO building, and the CHIME/FRB backend (hatched red) is in a third shipping container adjacent to the cylinders (see Fig.~\ref{fig:photo}). Note that the total input data rate into the F-Engine is 13 Tb/s.  The data rate into the CHIME/FRB backend is \fix{142 Gb/s.}}
    
    \label{fig:schematic}
\end{figure}

A schematic diagram of the CHIME telescope signal path is shown in Figure~\ref{fig:schematic} and further explained in the following sections. A total of 256 dual-polarization feeds, spaced by 30 cm, are placed along 80 m of the focal line of each of the four cylinders, giving a total of 2048 signal paths. 
Digital signal processing of these signals generates the multiple beams that make this telescope a powerful instrument for FRB research (see \S\ref{sec:xengine}).
The feed design \citep{mdc+14} achieves nearly equal 
beamwidths in both polarizations
and excellent matching over the octave operating band of the receiver.
Mutual coupling between adjacent and nearby feeds was considered in the
design of the baluns and matching networks. The entire feed is constructed from
printed-circuit materials, appropriate for mass production.

The focal line is deliberately kept simple, housing only feeds and low-noise amplifiers. Two low-noise amplifiers accept the signals from each feed, and those signals are carried through coaxial cables of equal length (50 m) to receiver huts (shipping containers that have been modified with radio frequency shielding and liquid-cooling).  One receiver hut serves two cylinders and is placed between them. Within each of the two receiver huts, signals from 1024 inputs are further amplified in a stage that includes a bandpass filter, \fix{and are digitized and split into 1024 frequency channels} by the custom F-Engine electronics (see \S\ref{sec:fengine}). Temperatures inside the receiver huts are controlled with a liquid-cooled system, but no attempt is made to control the temperature of focal-line components.



\section{Upgraded CHIME Correlator}
\label{sec:correlator}

The CHIME correlator was originally designed to handle the 2048 inputs from the CHIME antennas to map the CHIME-visible sky in redshifted 21-cm emission.  The CHIME correlator, of hybrid FX design, uses custom FPGA boards to digitize and channelize the data (the ``F-Engine''), while a GPU cluster provides the spatial correlation (the ``X-Engine''). The CHIME correlator was designed to be capable of recording visibilities across 400 MHz of bandwidth divided into 1024 frequency channels \fix{at $\sim$20-s cadence, sufficient for BAO mapping and radio frequency interference (RFI) excision.}

\begin{figure}[t]
	\includegraphics[width=\textwidth]{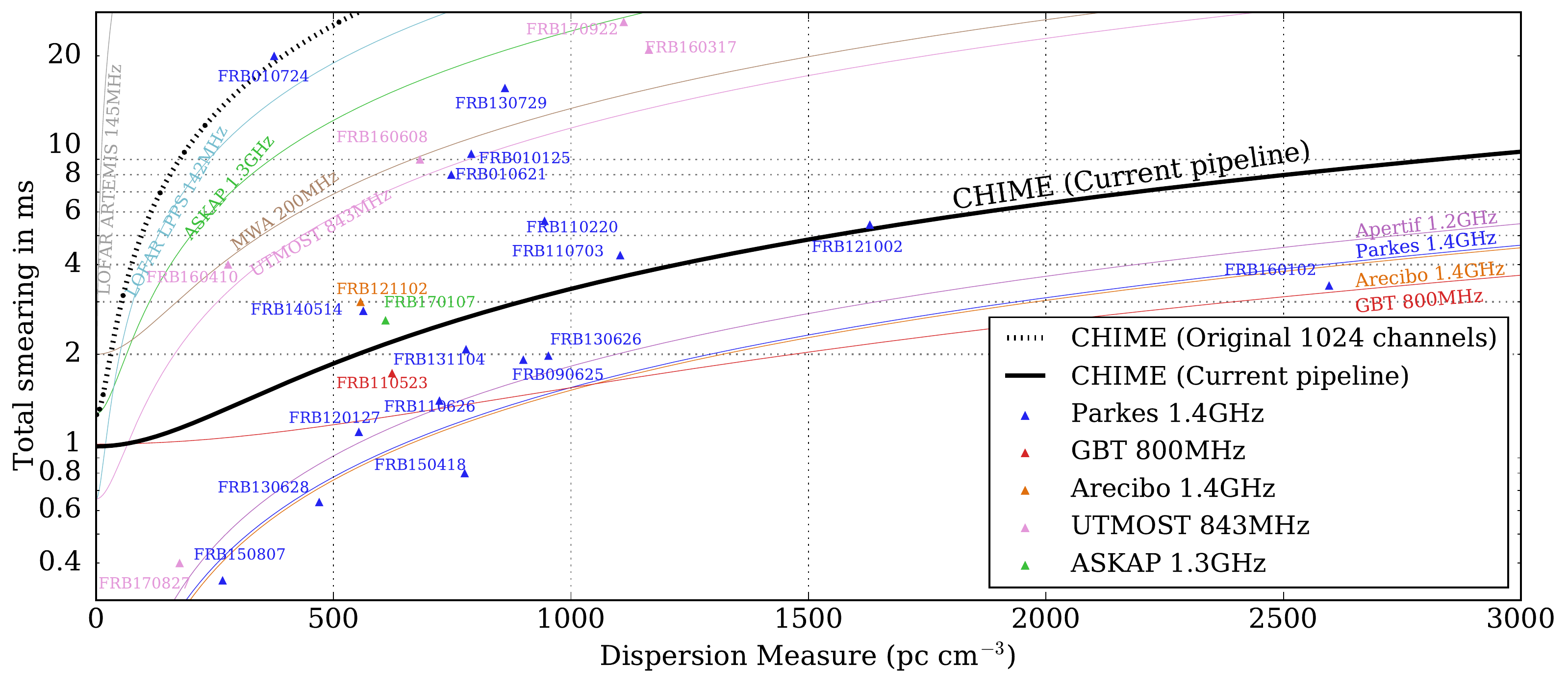}
    \caption{
Intra-channel dispersive smearing timescale as a function of DM for a variety of FRB searches including CHIME/FRB. The black dotted line would be the CHIME/FRB smearing if the native CHIME frequency resolution (1024 channels) were used.  Measured widths at 1.4~GHz of detected FRBs (color-coded by detection survey) are indicated.  Observed FRBs having widths far below the black dotted line would be artificially broadened to the black dotted line at the native resolution, rendering them difficult if not impossible to detect.  To mitigate this problem, upchannelization to 16k frequency channels is performed (see \S\ref{sec:beamforming}). The smearing for the current pipeline having 16k frequency channels is shown by the solid black line.  The colored lines are the intra-channel smearing of other FRB surveys, as labelled. }
    \label{fig:smearing}
\end{figure}

FRB detection, however, demands significantly higher time and frequency resolution, as well as spatially localized sky beams, albeit with less stringent calibration requirements. First, typical FRBs have durations of at most a few ms, so FRB surveys require at least comparable time resolution. Additionally, FRBs are dispersed by free electrons along the line of sight; without correction by dedispersion, their signals are rendered undetectable at the CHIME operating frequencies. Dedispersion of channelized intensity data, hence detection, demands high-frequency resolution, but leaves
a residual dispersive smearing within each frequency channel.
Minimizing this intrachannel smearing requires very narrow frequency channels, especially at CHIME's low operating frequencies. Figure~\ref{fig:smearing} shows the predicted intra-channel dispersive smearing time versus DM for a variety of FRB surveys. The CHIME/FRB project has opted for 1-ms cadence and 16384 (hereafter, 16k) frequency channels (see Table~\ref{ta:chime}) in order to minimize dispersion smearing in the most relevant part of phase space. 

The baseline CHIME correlator required upgrades to provide  the independent high-cadence, high-frequency-resolution and spatially discrete data streams necessary for FRB detection. The raw data output rate from the CHIME F-Engine is 6.5 Tb/s, making it difficult to duplicate or distribute beyond the X-Engine. The additional processing must therefore take place inside this system. The next section is a description of the correlator system, with emphasis on the modifications required for FRB detection.

\subsection{Digitizer, F-Engine, and Corner-turn}
\label{sec:fengine}

The custom electronics that perform the digitization, frequency channelization, and ``corner-turn'' (see below) are housed in two 20-ft steel shipping containers outfitted with radio-frequency shielded enclosures which provide $>$100 dB of shielding of the focal line from the
high speed electronics within. These enclosures (labeled the East and West receiver huts) are located midway between the 1st and 2nd cylinders and the 3rd and 4th cylinders, halfway along their lengths, to minimize and equalize the coaxial cable lengths that connect to the feed lines.

The F-Engine system consists of 128 ``ICE'' Motherboards \citep{bbc+16} housed in eight rack-mounted crates and interconnected with custom high-speed, full-mesh backplanes.
The system is described in detail by \citet{bcd+16}. Briefly, sixteen amplified and filtered analog sky signals are digitized in the second Nyquist zone on daughter cards attached to each motherboard at a data rate of 800 MHz with 8-bit accuracy. This information is transmitted to a Field Programmable Gate Array (FPGA) located on each motherboard. The total data rate digitized by the F-Engine is 13.1 Tb/s for the 2048 timestreams. The 400-MHz bandwidth of each timestream is channelized into 1024 frequency bins, each 390 kHz wide, using a polyphase filter bank. \fix{A programmable gain and phase offset are applied to each frequency channel and the data are rounded to 4+4 bit complex numbers, providing a total data rate of 6.5 Tb/s.}

\fix{The ``corner-turn'' 
modules in the F-Engine re-organize}
the channelized data from all the motherboards in a crate in order to concentrate the data
for a subset of frequencies into a single FPGA. A second 
corner-turn module
reorganizes the data between a pair of crates located in the same enclosure before offloading the data to the X-Engine GPU nodes (see \S\ref{sec:xengine}) through 1024 optical high speed transceivers using standard 10 Gb/s ethernet protocol. Data from all eight ICE motherboard crates are assembled in each GPU node, forming the last stage of the corner-turn. At this point, data for four frequency bins originating from all 2048 digitizers are assembled in a single GPU node, ready to be spatially correlated.

\subsection{X-Engine}
\label{sec:xengine}

The X-Engine consists of 256 processing nodes, 
each receiving 25.6\,Gb/s of frequency-channelized data on 4$\times$10\,GbE ports.
An Intel Xeon E5-2620v3 CPU handles data transfer, using the Intel Data Plane Development Kit (DPDK\footnote{\url{dpdk.org}}) to reliably achieve high throughput. Each node has 64\,GB of DDR4 RAM, sufficient to buffer resampled data for up to 20\,s; an upgrade to at least 96\,GB (for a 31-s buffer) is planned for the near future (see Section~\ref{sec:baseband}). The nodes occupy custom 4U rack-mount chassis and have no persistent local storage, but instead, a local set of file servers boot the nodes over the network.

Each node processes four input frequency channels of $\sim\,$390~kHz bandwidth each. These are processed by two dual-chip AMD FirePro S9300x2 GPUs, with each chip operating independently to perform all spatial and in-band processing on a single frequency channel. The processed data, including the resampled stream for FRB searching,
are exported to the backends over a pair of GbE links. 

The X-Engine is entirely liquid-cooled, using direct-to-chip cooling on the CPUs and GPUs and coupling the coolant to the ambient outside air with a 3$\times$120\,mm radiator in the front of each node, without any active chilling. Each rack operates a sealed and independent coolant loop, which is coupled to externally circulating coolant through a CoolIT CHx40 liquid-handling unit. The external coolant exhausts heat to the outside air through a high-capacity dry cooler. The X-Engine is housed in two 40-ft shipping containers directly adjacent to the cylinders (see Fig.~\ref{fig:photo}). To limit self-generated interference, a custom-built Faraday cage in each container \fix{provides $>$100\,dB of shielding from 1\,MHz to 10\,GHz.}

\subsection{L0: Beamforming and Frequency Channelization}
\label{sec:beamforming}

In order to take full advantage of the large instantaneous FoV of CHIME while maintaining the full sensitivity, we have developed a hybrid beamforming pipeline (hereafter referred to as the `Level-0' or L0 process; see Fig.~\ref{fig:pipeline}) to be employed in the X-Engine correlator. Details of this pipeline are described in \citet{nvp+17}. In summary, we synthesize 256 formed beams via a Fast Fourier Transform (FFT) algorithm along the North-South direction. The formalism of FFT beamforming in the context of a radio interferometer can be found in, e.g., \citet{tz09} or \citet{msn+17}. 
\fix{We zero-fill the FFT by a factor of two and resample the result to improve spatial alignment of the 256 formed beams across the 400-MHz observing bandwidth.}The final N-S beams are tiled across a runtime-defined range of angles (nominally $\sim 110^{\circ}$), evenly spaced in $\sin\theta$ space, where $\theta$ is the zenith angle. In the East-West direction, we form four beams via exact phasing. These give a total of 1024 discrete formed static beams that are closely spaced (with the exact spacing a tunable parameter) and tile the entire primary beam continuously. 
\fix{To increase spectral resolution, 128 successive 2.56~$\mu$s voltage samples are collected and Fourier-transformed. The square of the magnitude of this spectrum is downsampled in frequency by a factor of 8.  Three successive downsampled transforms are averaged together, 
and the two orthogonal polarizations are summed, producing a Stokes-I data stream with a cadence of 0.983 ms and a spectral resolution of 24.4 kHz.}
The output data thus consist of 1024 \fix{total intensity beams,} with 16k frequency channels at $\sim 1$-ms cadence.  These data are scaled, offset, and packed into 8-bit integers for transmission.  In principle it is possible to perform coherent dedispersion to a selected DM on phase data in the X-Engine,
however this feature has yet to be implemented.

%
%
%

\section{CHIME/FRB Instrument and Software Pipeline }
\label{sec:chimefrb}

\begin{figure}[t]
	\includegraphics[width=\textwidth]{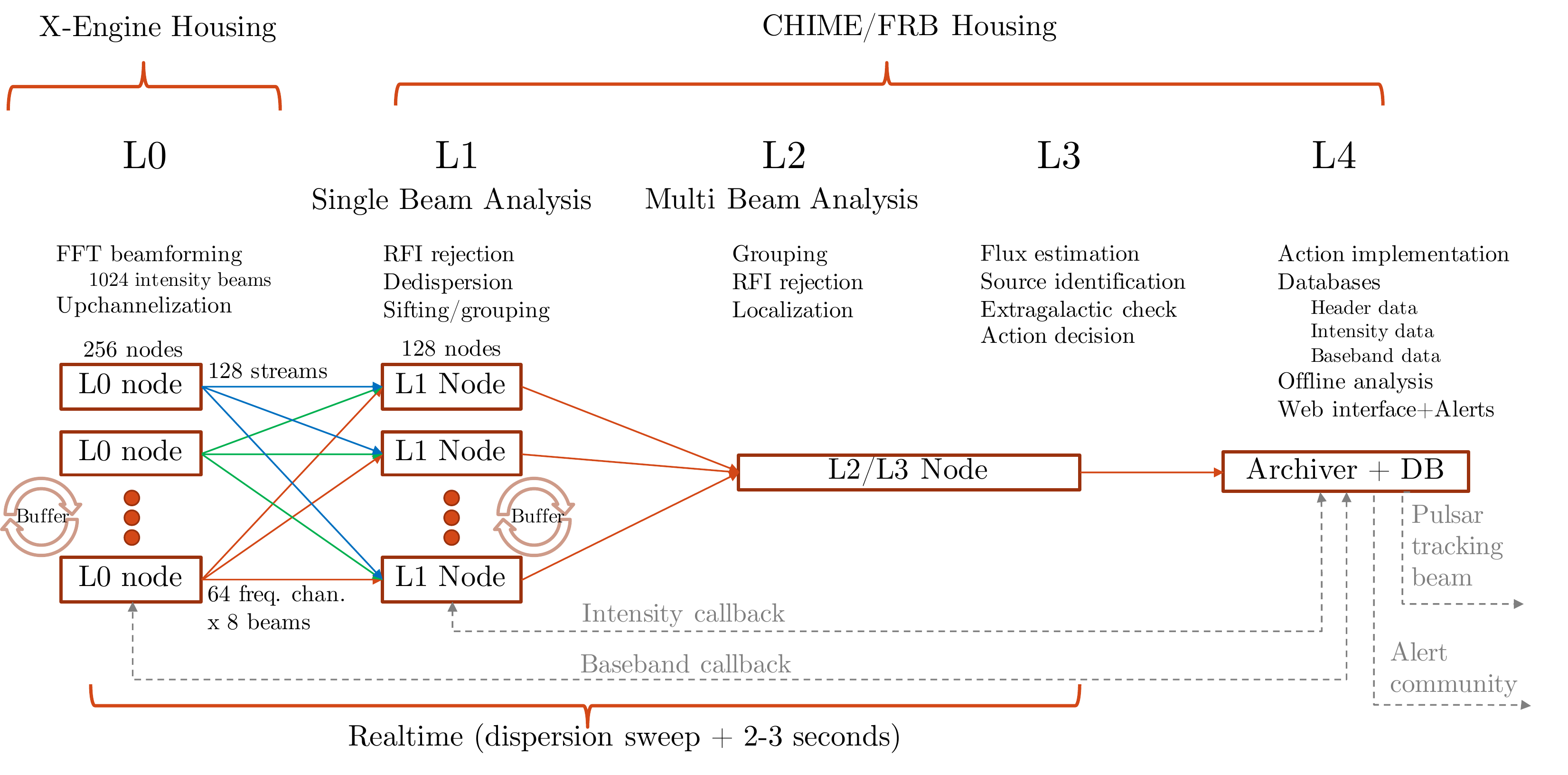}
    \caption{Schematic of the CHIME/FRB software pipeline. 
L0 performs the beamforming and up-channelization described in
\S\ref{sec:xengine} and \S\ref{sec:beamforming} with a limited bandwidth in each node.  
L1 performs RFI rejection and dedispersion described in \S\ref{sec:dedisp} using the full bandwidth for a limited number of beams.  Portions labeled
L2, L3, L4, described in \S\ref{sec:grouping} through \S\ref{sec:archiving}, combine information from multiple beams, classify the detections, provide alerts, and store events.  The L1 buffer, available in each node, is for intensity data call-back
(see \S\ref{sec:callback}) and the
L0 buffer is for the baseband call-back (see \S\ref{sec:baseband}).
}
    \label{fig:pipeline}
\end{figure}

The CHIME/FRB instrument is the system built to search for FRBs in real time, after receiving the 16k frequency channels at 1-ms cadence for the 1024 CHIME telescope beams from the upgraded X-Engine correlator.  
The pipeline is split into four further stages or levels, named L1 through L4.  
Figure~\ref{fig:pipeline} schematically represents the different components of the pipeline and the flow of data.  
Each correlator node calculates the intensities for four frequency channels for all beams and each L1 node runs FRB searches on the full frequency data for eight independent beams. Thus the data from the L0 nodes to the L1 nodes are cross-distributed across the clusters via the networking described in \S\ref{sec:network}.
L1 performs per-beam RFI rejection and dedispersion using a highly optimized tree algorithm \citep{s++18}, identifying candidate events in the DM/time plane.  L1 processing is performed on each of the 1024 formed beams by a dedicated cluster of 128 compute nodes, and candidate events are consolidated at L2.  L2 groups events seen simultaneously in different beams at the same DM and improves localization based on the strength of the signal in multiple beams.  L3 classifies the detection and selects among different actions---including alerting the community within a few seconds of the detection---based on source properties.  \fix{L4 performs the selected actions and hosts} a database that stores astrophysical events, including individual pulses from radio pulsars and RRATs, for further offline analysis.  
The L0 and L1 stages include buffers so that baseband (i.e. voltage) and intensity data, respectively, can be retrieved upon request and further analyzed offline when an event is detected.  

\fix{Below we provide details of the CHIME/FRB instrument and of each key processing step.  
The CHIME/FRB hardware is described in \S\ref{sec:instrument}, our network in \S\ref{sec:network}, and L1 though L4 in \S\ref{sec:dedisp} through \S\ref{sec:archiving}.
Retrieval of intensity data is described in \S\ref{sec:callback} and our RFI mitigation strategy in \S\ref{sec:rfi}. }


\subsection{CHIME/FRB Instrument}
\label{sec:instrument}

Similar to the X-Engine, the CHIME/FRB instrument consists of a total of 132 compute nodes housed in a customized RF-shielded 40-ft shipping container sitting adjacent to the telescope cylinders (Fig.~\ref{fig:photo}).  Each node contains two 10-core Intel E5-2630v4 CPUs and 128 GB of RAM in a 4U height (17.8~cm) case.  L1 processing, namely dedispersion (\S\ref{sec:dedisp}), is the most computationaly intensive, and runs on 128 of the nodes (see Fig.~\ref{fig:pipeline}).  Each L1 node processes eight beams using 16 cores.  The remaining four cores in each node are utilized for assembling the incoming
packets, processing the outgoing headers and intensity data retrieval (see \S\ref{sec:callback}).
L2 and L3 share a node, and L4 runs on a node of its own.  All nodes are network-booted for flexibility to allow dynamic
mapping of nodes to specific functions.  To alleviate single-point failures, the L2/L3 and L4 nodes each have a backup node that will automatically take over processing in the event of a node failure.  
For on-site data storage, a Storinator S45 Turbo unit (manufactured by 45 Drives Inc.) is fitted with 45 10-TB Western Digital Gold drives. The X-Engine and FRB systems are monitored and controlled through the Intelligent Platform Management Interface (IPMI)\footnote{E.g. \url{www.intel.com/content/www/us/en/servers/ipmi/ipmi-technical-resources.html}}. 

The FRB instrument is liquid-cooled, using the same direct
to chip cooling as for the X-Engine (\S\ref{sec:xengine}). However, as this system
does not have GPUs, there is far less heat loading per node
($\sim$200~W) compared to the X-Engine ($\sim$1000~W).  A smaller $240\times 120$-mm radiator couples the cooling fluid to forced air circulating in each node.
One CoolIT CHx40 liquid-handling unit is used to service two full racks that together contain a total of twenty nodes.


\begin{figure}[t]
	\centerline{\includegraphics[width=0.8\textwidth]{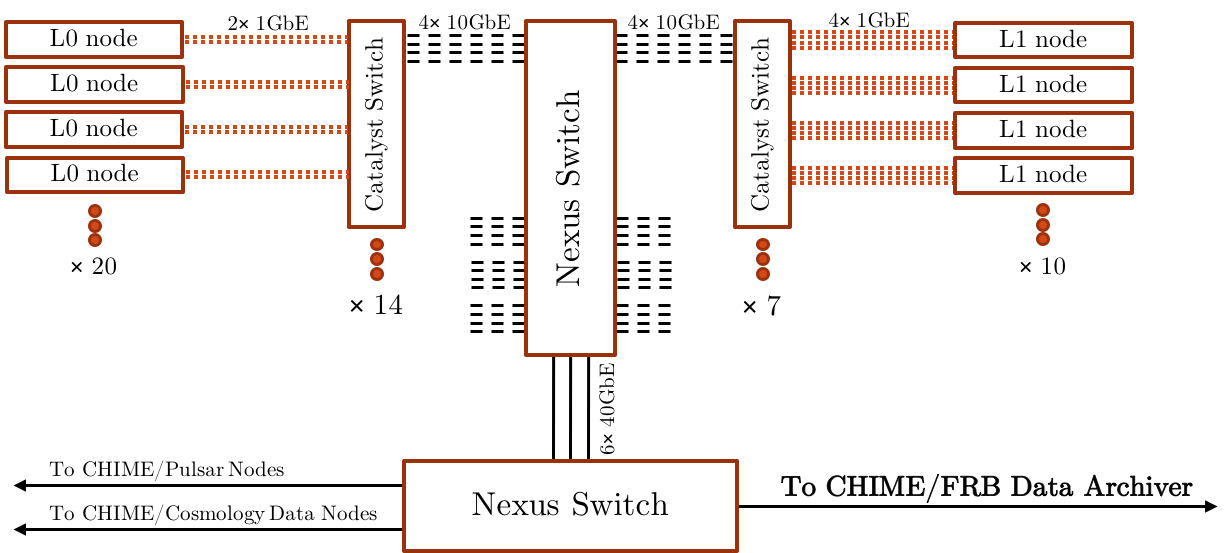}}
    \caption{A schematic of the CHIME/FRB data network. Each L0 node is connected via two 1~GbE connections (dotted orange lines) to a Catalyst switch which in turn is connected to the backbone Nexus switch via four 10 GbE links (dashed black lines). Similarly, each L1 node is connected via four 1~GbE connections to the Catalyst switches which connect to the Nexus switch. The two backbone Nexus switches are connected via six 40 GbE links (solid black lines). CHIME/Pulsar computation nodes, CHIME/Cosmology data storage nodes and CHIME/FRB data archivers are mounted on the network via the Nexus switches. 
}
    \label{fig:network}
\end{figure}

\subsection{CHIME/FRB Network}
\label{sec:network}

\fix{The X-Engine simultaneously transmits data to CHIME/Cosmology data-handling nodes, the FRB search backend and CHIME/Pulsar. For the FRB search backend, each X-Engine node processes 64 24.4-kHz frequency channels for all 1024 beams, while each L1 node processes all 16k frequencies for eight beams.  Thus, each L0 node must send data to each L1 node, which is achieved by employing three layers of network switches, as shown in Figure~\ref{fig:network}. Groups of twenty L0 nodes are connected, via two 1~Gb/s links, to a CISCO Catalyst switch (Model number 3650-48TQ-L) which routes the data to a central data switch (CISCO Nexus 3132Q-X) via 4~$\times$~10 GbE outputs. The data are then routed through a second set of Catalyst switches to groups of 10~L1 nodes on four  1~Gb/s links. In total, this network consists of 21 Catalyst (14 for L0 and 7 for L1) and two Nexus switches. One of the 1 Gb/s link on L0 is dedicated to transmitting data to the FRB and Pulsar backends, at a rate of 0.55~Gb/s and 0.25~Gb/s, respectively, with the transmission of each data packet timed to avoid data loss of the UDP packets. Each L1 node receives data at a rate of 1.1~Gb/s for a total data rate for the FRB system of 142~Gb/s. The CHIME/Cosmology data and FRB baseband data (when requested --- see \S\ref{sec:baseband}), are transmitted on the second L0 1 Gb/s link to archive machines.   The network traffic is continuously monitored for missing or corrupted data packets so that these effects can be included in the sensitivity calculations for the FRB search (\S\ref{sec:calibration}).}

\subsection{L1: Dedispersion and Candidate Event Identification} 
\label{sec:dedisp}

The most computationally expensive part of the CHIME/FRB search is 
the  ``dedispersion transform,'' 
which converts intensity data from time and frequency into time and DM, allowing efficient detection of dispersed impulse signals
(see Fig.~\ref{fig:pipeline}).
In our implementation of the dedispersion transform,
there are five key FRB parameters:  DM and arrival time (each of which range over a large number of trial values), 
spectral index, scattering time, and intrinsic width (which range over a
few trial values).
Thus the output of the dedispersion transform is a 5D array of 
signal-to-noise ratios (SNRs), to which we apply a \fix{tunable threshold, currently set at
$10 \sigma$,} to identify candidate events for processing
by subsequent stages of the pipeline.

The computational challenge of the CHIME/FRB search is immense: the
input data rate is 1.5 PB/day, and the dedispersion transform computes
$10^{11}$ SNR values per second (total for all beams).
This is orders of magnitude
larger than any FRB search to date, and comparable to the future search
planned for the Square Kilometer Array\footnote{\url{www.skatelescope.org}}.  For example, the data rate of the 36-beam ASKAP ``fly's-eye''
search~\citep{bsm+17} is 1700 times smaller than for CHIME/FRB.

To meet these challenges, we have developed a new FRB search code
``{\tt bonsai}," which we will release publicly in the near future.
This code will be described in detail in a forthcoming paper~(Smith et al. in prep.). \nocite{s++18} 
Here, we briefly summarize its main attributes.

There are two general approaches to dedispersion:
direct algorithms~\citep[see, e.g.,][]{mt77}, and
tree algorithms~\citep{tay74,zo17}. 
The computational cost of tree dedispersion is ${\mathcal O}(TF \log F)$,
versus ${\mathcal O}(TF^2)$ for direct dedispersion, where $T$ and $F$
are the number of time and frequency samples, respectively.
Thus tree dedispersion is parametrically faster, but suffers from
two problems in practice.
First, tree dedispersion tends to perform badly due to memory bandwidth bottlenecks.  
In a straightforward implementation, a large memory region (larger than the Level 3 cache available in the CPU of our nodes)
must be processed in each of $\log_2(F)$ iterations of the algorithm.  The cost of 
these memory transactions ends up being much larger than the computational cost.  Second, in 
some implementations, tree dedispersion may make approximations to the $\nu^{-2}$ 
dispersion delay which results in loss of SNR.

Our implementation of tree dedispersion, {\tt bonsai}, includes two new features which
solve these problems.  To address memory bandwidth bottlenecks, we have implemented
a version of tree dedispersion which is ``blocked,'' in the sense that the data are processed in a small number of passes, in blocks having size tuned to that of the CPU cache.  This minimizes the number
of times the data go through the CPU Level 3 cache,
resulting in a factor $\sim$30 speed-up on
multi-core machines.  An interesting property of the blocked tree algorithm
is that it is also incremental: given sufficiently large SNR, the search triggers within a few
seconds, even for an FRB whose dispersion delay in the CHIME band is much larger.
This low latency is critical for the triggered baseband recording system (\S\ref{sec:baseband}),
since the X-engine can presently only buffer data for 20 sec, and this window must
include both the dedispersion delay of the FRB and the latency of the search.

Second, we have made tree dedispersion close to statistically optimal
through appropriate choices of internal weightings.  For CHIME/FRB, we choose parameters
which are a compromise between computational cost and statistical optimality.
We search to a maximum DM of 13,000 pc~cm$^{-3}$, maximum pulse width of 100 ms, with two trial spectral
indices.
After initial RFI removal (see \S\ref{sec:rfi}), the search consists of multiple dedispersion trees with different levels of time downsampling, 
to cover different parts of the (DM, pulse width) parameter space.
The search is $> 80\%$ optimal over most of this parameter space, occupying 0.75 cores/beam, and a memory footprint of 7 GB/beam.  This
more than suffices to meet the initial science goals of the CHIME/FRB search.


The {\tt bonsai} code produces SNR estimates for a grid of DMs and times for a discrete set of spectral indices, scattering times, and intrinsic width parameters.
As a detail, the output array of SNR estimates is ``coarse-grained": we divide the grid of trial arrival times and DMs into coarse cells, and take the maximum of all SNR values over each cell.
We set a threshold at 10$\sigma$ and isolate local maxima within a neighborhood of size $\Delta\mathrm{DM} \approx 10$~pc~cm$^{-3}$ and $\Delta t \approx 0.25$~s.  Once we have identified the peak SNR parameters, we generate a lightweight description of the coarse-grained candidate event, the
``L1 Header."  This includes the DM, time and coordinates of the trigger, an RFI rating (see \S\ref{sec:rfi}), and SNR values for slices along each parameter dimension (holding the remaining parameters fixed at their peak values). 

%

To validate the accuracy and sensitivity of our candidate identification, we have performed injection analyses and processed CHIME Pathfinder \citep{baa+15} acquisitions containing pulsations from PSRs B0329+54 and B0531+21. Compared to a pipeline based on RRATTrap~\citep{kkl+15} and PRESTO\footnote{\url{github.com/scottransom/presto}} single pulse search \citep[e.g.][]{ckj+17}, we find in initial analyses that the CHIME/FRB L1 pipeline has a lower rate of RFI false positives, and is more effective at finding faint pulses (see \S\ref{sec:commissioning}).

\subsection{L2: Grouping of Multi-Beam Events}
\label{sec:grouping}


Per-beam event detection reports from the L1 {\tt bonsai} dedispersion analysis (see
\S\ref{sec:dedisp}) are streamed to a single endpoint where
they are buffered. These reports consist of L1 headers or null results.
Once all beams have reported for a given data
block, the existence of multi-beam events is investigated by grouping
candidates in time, DM, and sky position. The cluster detection is performed
with a simplified implementation of the 
DBSCAN\footnote{Density-based spatial clustering of applications with noise.} 
algorithm~\citep{eks+96},
which has $\mathcal{O}(n\log n)$ complexity, where $n$ is the number of events above our threshold, and is capable of handling
large event rates without issue.  
For any event in a group, there exists another whose differences in time, DM, and sky position are all below dimension-specific thresholds.  For position, the threshold ensures that grouped events are spatially connected through beam adjacency. The DM and time thresholds reflect the uncertainties built into the \fix{coarse-grained event properties} (see \S\ref{sec:dedisp}). 
\fix{After cluster detection, 
the event is classified as either RFI (see \S\ref{sec:rfi}) or astrophysical, and the L2 header is formed, containing one or more L1 headers from the grouping. RFI events are sent straight to the L4 database (\S\ref{sec:archiving}) while astrophysical events are passed to L3, after position refinement and event flux estimation, described next.}



Position refinement is
performed by comparing the detected SNRs from each beam to what is expected
from a frequency-dependent beam model. 
\fix{For multi-beam events, the detected SNRs are compared to those in a
pre-computed look-up table containing relative SNRs for a grid of sky
positions and spectral indexes to form a refined estimate of these parameters.}
\fix{$\chi^2$ minimization is then used to refine the initial guess and estimate the uncertainty region.}  For
single-beam events, a precomputed mapping between event SNR and the location 
and uncertainty region is used. The mapping is constructed from the
distribution of simulated events drawn from the beam model that produce
non-detections in adjacent beams. 

\fix{With a refined position and estimate of the on-axis SNR,  
radio fluxes of events are then estimated in real time. }
The estimate is 
based on the radiometer equation for single pulses \citep[see e.g.][]{cm03}. Besides instrument and 
pipeline-specific parameters, we use the SNR and pulse width as estimated by \texttt{bonsai} as input. 
The SNR is corrected for the beam sensitivity at the location of the event and the receiver bandwidth is 
corrected for the masked fraction from RFI excision. The sky temperature at the location of the event is 
interpolated from the reprocessed Haslam 408-MHz sky temperature map \citep{rdb+15} and scaled to our survey 
central frequency (600 MHz) using a power-law index of $-$2.6. The intrinsic pulse width is estimated 
by correcting for the sampling time, intra-channel smearing and scattering, in quadrature. Intergalactic scattering 
times are estimated as in \S3.11 of \citet{ymw17}, where we estimate scattering within our Galaxy using the 
empirical power-law fit by \citet{kmn+15} and ignore potential scattering in the host galaxy of an FRB.

\subsection{L3:  Identification of Extragalactic Events and Action Determination}
\label{sec:extragalactic}


Astrophysical events can be of Galactic or extragalactic origin, and can come
from either a known or an unknown source. The processing-related actions
triggered for an event depend on its identification and, as detailed below,
the pipeline makes this decision in L3 based on the determined DM and localized position. By automatically
recognizing events that come from known sources, we can provide alerts
(see \S\ref{sec:alert}) only for FRBs, and not for Galactic objects.
However, all astrophysical L1 events are stored, and
additional data products for non-FRB events can also be stored for
selected interesting sources.

We maintain a database of known sources that
initially consists of all known pulsars, RRATs and FRBs and
their relevant parameters recorded in the ATNF pulsar
catalog\footnote{\url{www.atnf.csiro.au/people/pulsar/psrcat/}}
\citep{mhth05}, the
`RRATalog'\footnote{\url{astro.phys.wvu.edu/rratalog/}}
and the FRB catalog\footnote{\url{www.frbcat.org/}}
\citep{pbj+16}. As time progresses, discoveries by CHIME/FRB and
other FRB searches will be added to the database. For each astrophysical
event, we compare the sky position and
DM, weighted by the measurement error, to the same parameters of the
sources in the database. The likelihood ratio (i.e. the Bayes factor)
of association with neighboring known sources is calculated. A threshold
for association is determined based on the Receiver Operating Characteristic (ROC) curve (the true positive
versus the false positive rate) for a set of simulated events. Presently,
only the sky location and DM of an event are being compared with those
of the known sources, but the framework allows for future additions
(such as a comparison of the periodicity or the flux).

For events that cannot be associated with a known source, the predicted
maximum Galactic DM along the line-of-sight is calculated
using models described by \citet{cl01} and
\citet{ymw17} of the Galactic free electron density. 
The difference in predicted maximum Galactic DM between the two models is taken as a systematic uncertainty; this is then added in quadrature with the L1-estimated DM
uncertainty to yield a full uncertainty, $\sigma$.  We impose the condition that a $\sigma$ less than 
a minimum value (currently set to 20\%) is set to this minimum, to avoid underestimating the true uncertainty when the two models
agree coincidentally. 
In the real-time pipeline, a source is deemed
extragalactic if its measured DM exceeds the maximum Galactic DM predicted
by \textit{both} models by at least 5$\sigma$. Sources with a measured DM
only 2$\sigma$ to 5$\sigma$ in excess of the predicted maximum Galactic DM fall
in an ambiguous class of sources. This class is invoked to identify
sources that may be Galactic, warranting multi-wavelength study to identify
possible Galactic foregrounds such as H\textsc{ii} regions.  Alternatively, they may
be extragalactic, but relatively nearby, hence potentially good targets
for multi-wavelength counterpart identification. 

Finally, a set of predefined rules determine which actions to trigger for
each detection.  Possible actions are to ignore an event,
store the event header in the events database (the default action), call
back for buffered intensity data from L1 (see \S\ref{sec:callback}),
dump buffered baseband data from L0 (see \S\ref{sec:baseband}), ask the
CHIME/Pulsar observing system to track the source position in search
mode, or send out an alert (\S\ref{sec:alert}).  The rules can be set
for individual sources or for groups of sources, with all rules being
stored in a configuration database that is accessible through a web
interface. Some examples of action rules are: ``if the flux of a binary
millisecond pulsar pulse is one 100 times higher than its nominal flux, store
the event header and trigger an intensity call-back" and ``if the source
is unknown and the event extragalactic, store the event header, trigger
an intensity call-back, a baseband dump and alert the community."


\subsection{L4:  Action Implementation and Event Archiving}
\label{sec:archiving}

\fix{L4 first implements the actions selected in L3 for each event, such as requests
to L1 for intensity data call-backs, requests for monitoring with CHIME/Pulsar,
or sending out real-time alerts to the community for FRB detections.
L4 is also host to the CHIME/FRB archive,} which saves the header information and its associated analysis products in a relational database for every event sent past the L1 stage regardless of whether it is classified as astrophysical or RFI in the L2 stage. Each event header that arrives at the L4 stage of the pipeline is assigned a unique event number to identify the event and track all data products associated with it.


The size and quantity of data products stored for each event depend on the actions and priorities determined at the previous stages of the pipeline. \fix{Due to the current limited bandwidth out of the observatory, a local archiver with 450 TB of disk space has been installed. The data will be regularly transferred to a long-term offsite staging facility.}
In order to manage file transfers between different locations, we employ a custom file management software \texttt{alpenhorn}\footnote{\url{github.com/radiocosmology/alpenhorn}} \citep{hsc+14}. \texttt{Alpenhorn} ensures reliable data transfer between multiple originating, intermediate, and archival storage sites. Among its features are keeping at least two copies of the data at different sites at all times, verifying data integrity through checksums, and handling data transfer among sites. The FRB pipeline database is periodically updated with the current status of the data files tracked by \texttt{alpenhorn}, so that a user accessing the live website has access to their location.

The database is integrated into a web display that allows for searching and filtering of events, displaying the best data measurements, interacting with data plots, tracing the location and requesting downloads of raw data. The web display also incorporates user accounts, allowing users to rank, classify, and comment on candidate events. Users can also update action criteria and known source properties.

\subsection{Intensity-Data Call-back}
\label{sec:callback}

Since the data rate of CHIME/FRB is very high, we cannot save to disk all the  intensity data taken by the telescope. Instead, each L1 node stores a buffer of recent data that can be retrieved if an interesting event is detected by L3 (see \S\ref{sec:extragalactic}). We save data from the beams in which the event is detected independently as well as the immediately adjacent beams. 

CHIME/FRB searches up to a DM of 13,000~pc~cm$^{-3}$.  At this maximal DM, a pulse can take up $\sim 250$~s to sweep through the CHIME frequency band.  The L1 nodes do not have enough memory to store this much intensity data at full resolution, so we implemented a telescoping ring buffer:  older data are progressively downsampled and stored with coarser time resolution to optimize memory usage.
Initially, we keep 60~s of the most recent data at full resolution, 120~s binned by a factor of two, and 240~s binned by a factor of four. 

As the L1 nodes do not have local storage, \fix{called-back intensity data are written to a network-shared archiver (see \S\ref{sec:archiving})} where they are available for additional offline analysis and visualization. Apart from generating the dynamic spectra (``waterfall plots'') for visualization of single pulse events, we refine the event parameters (DM, arrival time, pulse width) determined by the L1 pipeline. We perform an improved localization using the intensity data from neighboring beams (considering detections as well as non-detections) and obtain more accurate values for spectral index and flux. 

\subsection{RFI Excision}
\label{sec:rfi}
The CHIME/FRB software pipeline mitigates RFI using different criteria at different stages in order to minimize the misclassification of terrestrial signals as astrophysical transients. 

In L1, we remove RFI signals from the intensity data by iteratively detrending and subtracting a time series obtained by assuming zero DM \citep{ekl09} from the intensity data, while clipping amplitude and standard-deviation outliers. We have developed fast C++/assembly kernels which form a sequence of ``transforms'' at a high-level.  Each transform operates on arrays of intensity data with a parallel array of weights (where masked intensities are represented as zeros). The output of each transform is the input to the next. We have utilized low-level ring buffering, and transparent resizing and resampling of the data, since the transforms need not use the same block size or downsampling. 

Our current default RFI removal scheme consists of a sequence of nine clipping
transforms (five with 3$\sigma$ threshold and four with 5$\sigma$ threshold), which are iterated six times. The sequence is repeated twice; after each
iteration the intensity is detrended by two polynomial (degree 4) and spline (equivalent degree 12)
transforms, one along the time axis and one along frequency, respectively.
We have found empirically that this transform chain removes RFI in data we
have captured so far (e.g., from the CHIME Pathfinder, the Galt 26-m telescope, or with a CHIME incoherent beam -- see \S\ref{sec:commissioning}) 
to a sufficient level that simulated FRBs near detection
threshold dominate the output of the dedispersion transform.
Our real-time detrending and RFI excision pipeline is currently in a proof-of-concept
stage and we are still experimenting with the implementation.

After candidate events have been identified (see \S\ref{sec:dedisp}), and before they are forwarded to L2 (see \S\ref{sec:grouping}), a second round of RFI excision is performed by subjecting the SNR behavior surrounding significant detections to a machine learning classifier --- in this case, a support vector machine
\citep{cv95a}. Presently we include two features with the goal of capturing both local and global SNR behaviors. For the former, we use the SNR fall-off at nearby but sub-optimal DMs. For the latter, we consider the distribution of other above-threshold locations in the DM-time plane, noting that RFI tends to manifest as vertical streaks of events covering a large range of DMs.  As our chosen machine learning method
requires supervision, we have developed tools to allow for efficient event
labelling; training the classifier and refining feature definitions will be an
ongoing process throughout commissioning.

The candidate events detected in single beams are transmitted from all L1 nodes
to the L2 node. At this stage we group events detected in spatially neighboring beams and within a DM and time threshold. We further classify these events as astrophysical or RFI based on the distribution of SNRs in neighboring beams and the shape of the grouped beams. Astrophysical events are in the far-field of the telescope ($\gtrsim$ 25--50\,km for CHIME's operating frequency band) and are expected to have a sharply focused intensity distribution pattern whereas RFI events are typically in the near-field and likely detected in the horizon-directed sidelobes causing a significantly broader intensity distribution pattern.

\fix{In L2,} we use a machine learning classifier,
implemented using \texttt{scikit-learn}\footnote{\url{www.scikit-learn.org}} \citep{pvg+11}, to calculate the
probability of an event being astrophysical or RFI. 
Ultimately, multiple different classifiers can be loaded and
their calculated probabilities are assumed to be independent by the framework and multiplied together. Currently, the classifier is based on a stochastic gradient descent classifier with a linear combination of the following features: (1) the pattern of beams in a $3\times3$ grid in which the event was detected above threshold with respect to the beam with the highest SNR, (2) the ratio of the second highest SNR to the highest SNR in the group, (3) the ratio of the average SNR of the detected beams
to the highest SNR in the group, (4) the total number of beams where the event
was detected, and (5) the highest L1 grade associated with the group from the
L1 classifier. For events detected in single beams, no extra information
is available, so the grade from the L1 classifier is the only value used for
classification. In the future, we may implement features based on (1) the total
L1 event activity (number of candidates detected in the same time window) at
the time of event detection, (2) ratio of the FFT-formed beam intensity to that
of an incoherent beam formed from all CHIME feeds, and (3) location of the beams on the
sky (as RFI seems more likely to be detected at lower elevation beams).

As with L1, we are developing event labeling and generation of training sets
for the classifier. At the L2 stage, all L1 candidates that are sent are
stored in the database irrespective of their RFI/astrophysical classification.
This allows us to reclassify and retrain the classifiers as required in the future.

\section{Planned CHIME/FRB Features}
\label{sec:planned}

Here we describe features of the CHIME/FRB analysis pipeline that are either currently being developed or
are still in a design phase.

\subsection{Real-time Alert System}
\label{sec:alert}

Rapid reporting of FRB discoveries is important for coordination of detections with FRB searches at other radio bands, for finding repeat bursts from FRBs and for discovering or ruling out possible electromagnetic or gravitational wave counterparts. Specifically for CHIME's FoV, facilities such as the Owens Valley Long Wavelength Array (23--88\,MHz, Hallinan et al., in prep.) or the DSA-10 (1.4\,GHz, Ravi et al. in prep.) could provide constraints on the spectral indices, spectral ranges, absorption and scattering properties of FRBs. The non-Poissonian clustering of FRB\,121102 \citep{ssh+16b} makes the detection of repeat bursts more likely after the initial detection. If this characteristic holds for other FRBs as well, the monitoring of CHIME/FRB-discovered sources with tracking telescopes would be an important test of repetition as well as an opportunity for interferometric localization (e.g. with the Jansky Very Large Array\footnote{\url{www.vla.nrao.edu/}} in New Mexico as in Chatterjee et al. 2017).

The CHIME/FRB real-time alert system will follow the \texttt{VOEvent} implementation suggested by \citet{phb+17} to be compatible with the other FRB search efforts around the world. After a period of human confirmation at the beginning of the CHIME/FRB program, we plan to announce every detected FRB along with the significance of the detection and other relevant characteristics in real time. The latency of the CHIME/FRB real-time pipeline from the arrival of the lowest frequency signal is $\sim$ 2--3\,s. FRBs for which the intensity data are stored and re-analyzed (see \S\ref{sec:callback}), or for which there has been a baseband dump and subsequent improved parameter determination (see \S\ref{sec:baseband}), will have updates to their reported characteristics announced as soon as possible.

\subsection{Calibration, Sensitivity, \& Completeness Measurement}
\label{sec:calibration}

CHIME/FRB is a complex instrument with multiple components that must be regularly calibrated and monitored to reach its science goals. 
\fix{Here we describe the three planned stages of the process: phase calibration of CHIME, monitoring of the pipeline performance to track
CHIME/FRB's exposure to the sky, and measurement of the sensitivity and completeness of the CHIME/FRB pipeline.}

\fix{In order to form phased beams, each RF input signal must be phase-calibrated regularly.
While ultimately milli-radian phase calibration} will be available for CHIME in order to meet the stringent requirements of the cosmology experiment, this calibration accuracy will not be available at the commencement of FRB searching. For the CHIME/FRB system, we will determine the complex gains in a near real-time system using transits of bright effective point sources like Cassiopeia A and Cygnus A \citep{tre+17}, approximately once per day.
We will observe such a calibration source with a full set of $N(N-1)/2$ visibilities, where $N$ is the number of feeds, using the CHIME cosmology X-Engine at a full complement of radio frequencies. 
We will then compensate for the delay expected from the physical locations of the antennas and the calibrator sky location. Any observed residual delay in the response to the calibrator will be instrumental, which we record and compensate for in the FRB beamformer in L0. The observations of the calibrator sources will also be used to calibrate the CHIME bandpass.

Most scientific goals of CHIME/FRB --- measurements of the FRB rate and of property distributions --- require a thorough understanding of CHIME/FRB's detection sensitivity as a function of time and sky-location (i.e. the exposure of the survey) and the completeness of its survey as a function of burst fluence, burst width, DM, etc. 
We take two approaches that allow us to cross-check our results. To determine the exposure and the sensitivity as a function of time, we have to monitor metrics from the data flow and compute a sensitivity as a function of time and beam location using the radiometer equation \citep{mc03}. However, the multiple complex stages of RFI filtering and excision make it challenging to understand and study the completeness, especially in the presence of rapidly changing RFI that is correlated across phased beams. Hence we also plan to create a real-time multi-beam FRB injection and detection system to estimate the survey completeness in real time. 

To account for the total sensitivity and exposure of CHIME/FRB, we have to keep track of the sensitivity of each coherent beam and the number of beams that are being searched for FRBs at a given time. At the L0 stage, CHIME/FRB will monitor which feeds from the telescope are usable (informing beam shape and sensitivity) and which L0 nodes are active (informing the bandwidth and usable frequency channels).  At the L1 stage, we monitor the UDP packet loss from L0 nodes to the L1 nodes (again informing usable bandwidth and frequencies), the fraction of data masked by the first stage RFI removal algorithm, and the number of L1 nodes that are actively searching for FRBs, informing the sensitivity of each beam on the sky. At the L2/L3 stage, the number and identity of the reporting L1 nodes, and the delay in TCP packets are tracked.

Most of these metrics will be reported on a few-minute timescale and stored in a time-tagged database. A database scraping script will then collect the metrics and calculate the sensitivity and exposure of CHIME/FRB to the sky.

\begin{figure}[t]
	\includegraphics[width=\textwidth]{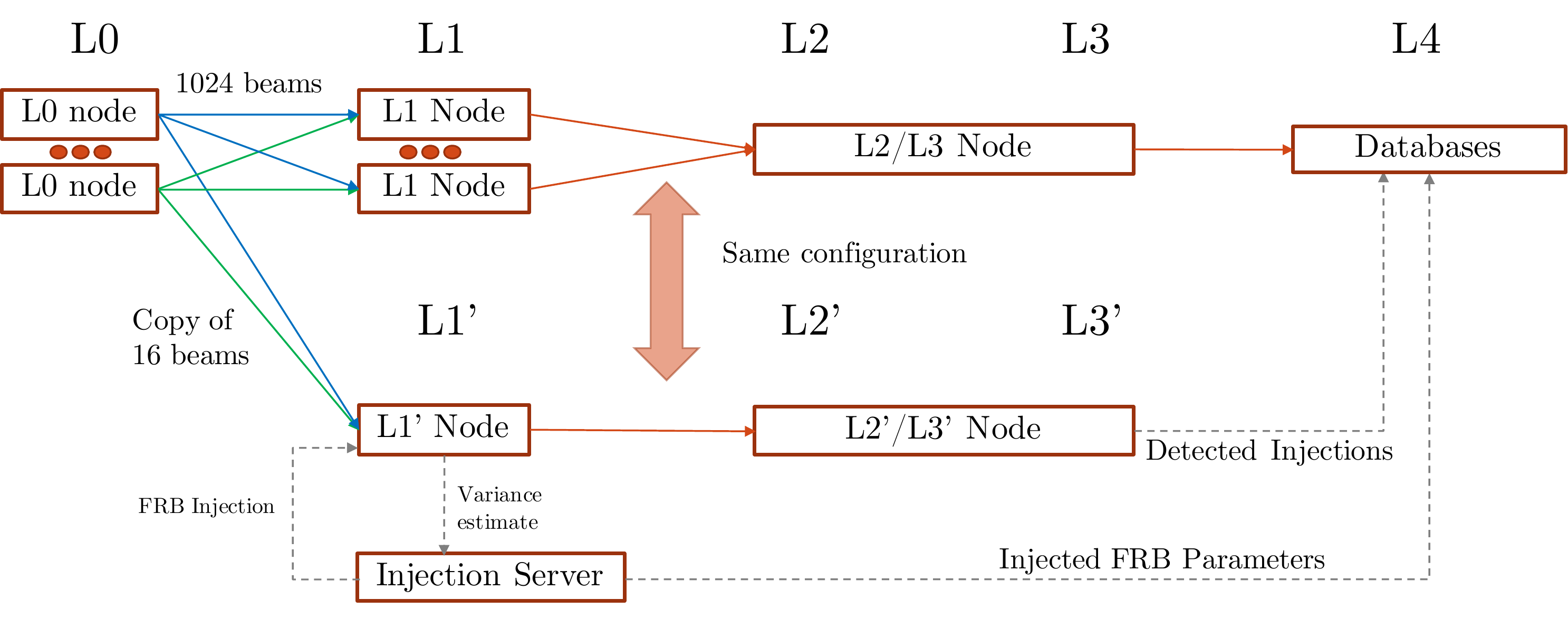}
    \caption{Schematic of the planned CHIME/FRB injection pipeline. A fraction of the CHIME/FRB beams (typically a 4$\times$4 grid) is copied to a separate L1 node where FRB signals with a large range of fluence, DM, width and scattering, (convolved with modeled beam responses) are injected into the beams. The processing of the injected data will be identical to that in the main CHIME/FRB pipeline and the detections and non-detections of the injected FRB signals will be tracked in a database. }
    \label{fig:injection}
\end{figure}

\fix{In order to test the response of our RFI excision and filtering, we plan to inject FRB signals of known properties, accounting for the spatial and spectral responses of the coherent beams, into a small number of the beams and verify we recover them as expected.}
Figure~\ref{fig:injection} shows a schematic of the injection pipeline that will run in parallel to the main CHIME/FRB search pipeline. A grid of 4 $\times$ 4 coherent beams will be chosen randomly among the 1024 beams to sample different North-South locations in the primary beam. An injection server will generate a set of FRB parameters (location, flux, DM, width, scattering, and spectral index) from a prior distribution. The effective signal strength in each beam and frequency band will then be computed  based on the coherent beam model. A simulator code will inject the signal into the intensity data from the corresponding beams. The rest of the data analysis will proceed identically to the CHIME/FRB search pipeline. The injected parameters and the recovered parameters will be stored in a temporary database where the detection completeness of the pipeline as a function of FRB parameters can be analyzed. 
\fix{A detailed paper reporting the measured sensitivity of the CHIME/FRB search pipeline will be forthcoming once commissioning (\S\ref{sec:commissioning}) is complete.}


\subsection{Planned Triggered Baseband Recording System}
\label{sec:baseband}

We plan to have triggered recordings of baseband data containing FRBs. This system will provide CHIME/FRB with sensitivity to polarization, greatly improved ability to localize sources, spectral and time resolution constrained only by the uncertainty principle $\Delta f \Delta t \geq 1$, and opens possibilities to perform very-long-baseline interferometry (VLBI) of FRBs with other telescopes. This baseband system is presently under development and will be commissioned once CHIME/FRB science operations are underway. Here we provide our current plans for this system.

Baseband data from all 2048 correlator inputs that have been spectrally channelized by the F-Engine will be buffered in the X-Engine prior to beamforming. These data will be written to disk on an archiver system upon receipt of a trigger from the FRB search pipeline (see \S\ref{sec:extragalactic}). Ultimately, this system will allot approximately \fix{90\,GB} of random access memory in each X-Engine node providing a 31-s ring buffer --- long enough for the few seconds of latency in the FRB search pipeline as well as the dispersion delay across the band for DMs up to $\sim 1500\,\textrm{pc}\,\textrm{cm}^{-3}$. To economize network resources and disk space, only the 100\,ms of data surrounding the burst will normally be read out and stored to disk, for each frequency channel. These data will be analyzed offline in the initial implementations of the system. The buffer system will be designed so that upon receipt of a trigger only a small block of memory containing the desired 100\,ms of data will be frozen, with the rest of the memory continuing to participate in the ring buffer. As such, the system will have no down time during readout of baseband data, only a slight shortening of the buffer.

Among the most compelling applications for the baseband system is improved localization precision of FRB sources. Rather than sources being localized to a formed beam width of $\sim \lambda/D$ (where $\lambda$ is the observing wavelength and $D$ is the telescope extent), the baseband system permits interferometric localizations with precision \citep{msn+17}
\begin{equation}
  \Delta\theta = \frac{\sqrt{6}}{2\pi} \frac{\lambda}{D}\frac{1}{\rm SNR}\;.
\end{equation}

Here SNR is the signal-to-noise ratio of the burst as detected in the baseband data and the geometric factor of $\sqrt{6}$ accounts for CHIME's baseline distribution. This will amount to arcminute uncertainties for most bursts. 

The baseband system, through the use of coherent dedispersion \citep{hr75}, will also permit the identification of structure in the FRB signals at time resolution well under the 1-ms provided to the
CHIME/FRB search engine.  Any such structure will constrain FRB emission physics. Similarly, the baseband data will allow the detailed investigation of FRB scattering and scintillation timescales.  These properties can be used to constrain the astrophysical environments in which FRBs occur \citep[e.g.][]{mls+15a}. 

Finally, baseband data will contain full Stokes information, allowing us to determine the degree to which each FRB is polarized  \citep{pbb+15,mls+15a,rsb+16,msh+18}.  In cases where an FRB has a linearly polarized component, the dependence of the polarization angle on observing frequency can then be used to calculate the Faraday rotation measure (RM). The RM corresponds to the integral of the electron-density weighted line-of-sight magnetic field between the observer and the FRB, and thus is a powerful probe of an FRB's immediate environment, host galaxy and the intervening intergalactic medium \citep{pkb+15,arg16,rsb+16,msh+18}.  

\section{CHIME/FRB Commissioning}
\label{sec:commissioning}

\begin{figure}[t]
	\includegraphics[width=0.6\textwidth,clip=True,trim=0.1in 0.1in 0.15in 0.1in]{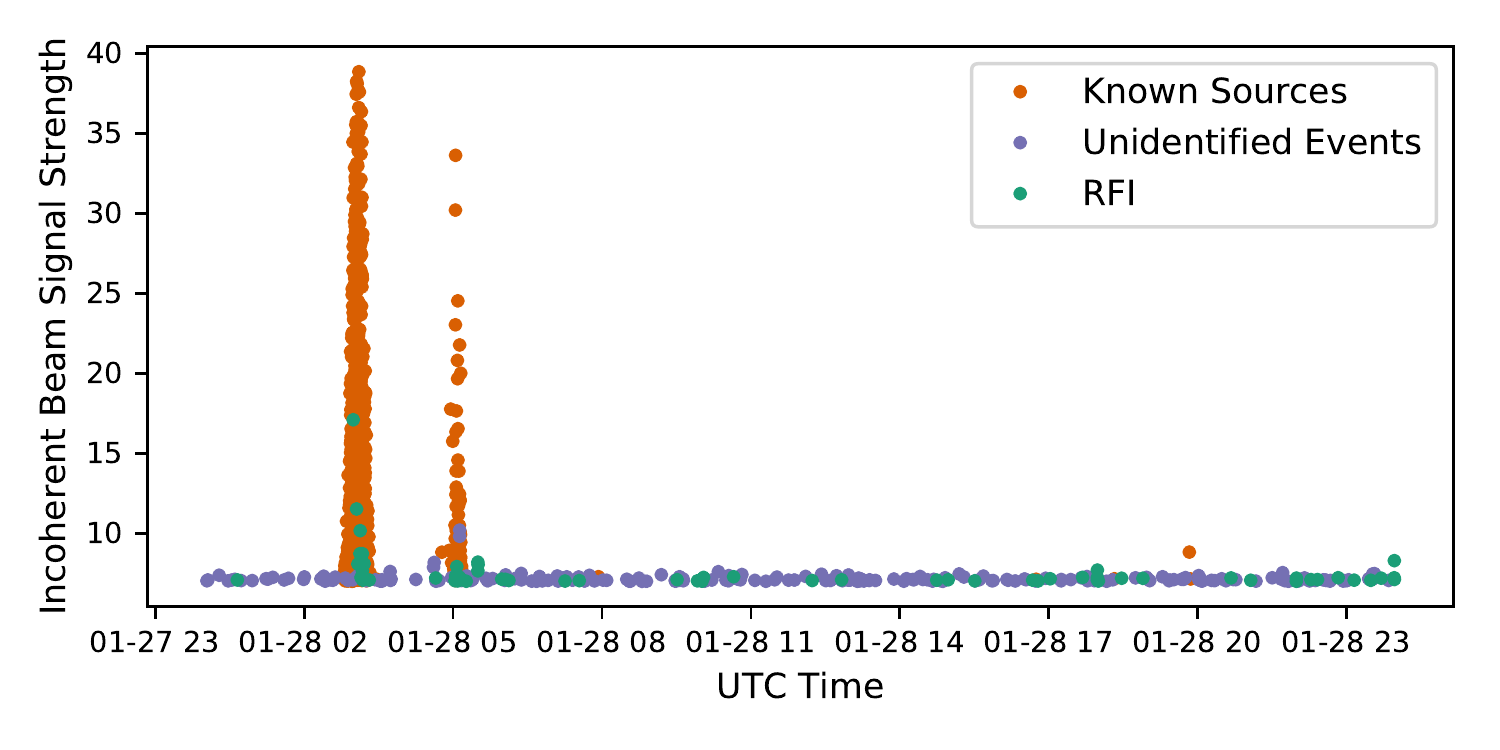}
    \includegraphics[width=0.4
    \textwidth,clip=True]{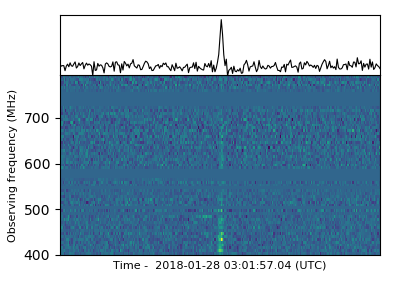}
    \caption{CHIME/FRB incoherent beam commissioning and test observations of transits of PSR B0329+54 and the Crab pulsar (orange dots), obtained on 2018 January 28 and 29.  (Left) Approximate incoherent beam signal strength (greatly reduced compared to the expected coherent beam signal strength once available) versus time over 24 hrs, showing one transit of both pulsars
    (PSR B0329+54 on the left and the Crab pulsar to its right). The x-axis label denotes UTC date and time. (Right) Waterfall plot of one bright single pulse from PSR B0329+54 observed with the CHIME/FRB 10-node system on 2018 January 28.  }
    \label{fig:0329}
\end{figure}

The CHIME/FRB pipeline development has followed a phased approach. First releases
of software were developed on a two-node L1 system at McGill University
in early 2017. A ten-node system was constructed at UBC to allow
for testing before this system was moved
to the CHIME site in late 2017. The deployment of the ten-node
system has allowed initial integration with the CHIME X-Engine and L0 system, as well as testing
of the backbone network and pipeline configurations. Initial testing of
the beam-forming and phase calibration is underway with the CHIME
telescope.  Observations with an incoherent beam (summing the intensities of all feeds without consideration of their phases) for limited times have been done.  Single pulses from multiple different radio pulsars have been detected at approximately the expected signal strength, noting the greatly reduced sensitivity compared with that expected for coherently formed beams.
Figure~\ref{fig:0329} (left) shows a synopsis of observations from 2018 January 28 showing the transits of PSR B0329+54 and the Crab pulsar (see figure caption for details) and Figure ~\ref{fig:0329} (right) shows a dynamic spectrum/waterfall plot for one bright pulse from PSR B0329+54.
The detection of single pulses from several radio pulsars with the 10-node system has allowed an initial commissioning with 64 beams on the
sky, as well as verification of basic system operations, including time-tagging, event clustering and identification and archiving. The full system node-level components are being integrated at 
McGill University in early 2018 prior to deployment at the CHIME site in Spring 2018. Science operations are expected to commence in the latter half of 2018.

\section{Predicted CHIME FRB Event Rate and Envisioned FRB Science}
\label{sec:rate}


FRBs have been detected mostly at 1.4 GHz \citep[e.g.][]{tsb+13,sch+14}, with some near 800 MHz \citep{mls+15a,cfb+17}, however thus far none has been seen below 400 MHz in spite of significant exposures with large telescopes at those frequencies \citep[e.g.][]{kca+15,rbm+16,ckj+17}. The reasons for the apparent low-frequency dearth are not yet understood, which makes CHIME's band particularly important for making progress on FRBs. \citet{abb+17} reported on an FRB search in the 400--800-MHz band, but this was using the CHIME Pathfinder telescope in an incoherent mode, which would have been able to detect only extremely bright FRBs. CHIME/FRB will have several orders of magnitude more sensitivity than that study.

There are three main reasons why the FRB rate in the CHIME band may be significantly lower than at higher radio frequencies.  The first is that multi-path scattering of radio waves by inhomogeneities in interstellar, or in principle intergalactic, plasma causes pulses to be broadened significantly.  This is strongly dependent on radio frequency $\nu$, with broadening time varying as $\nu^{-4.4}$ \citep[e.g.][]{ric77}.  It is possible that some FRBs are broadened beyond detectability in the CHIME band, particularly at the lower end.  However, many observed 1.4-GHz FRB widths are below 1~ms (see Fig.~\ref{fig:smearing}) and still are limited by intra-channel dispersion smearing,
suggesting scatter-broadening may not be a hindrance to detection in the CHIME band, especially given that we plan to search to widths of up to 100 ms (see \S\ref{sec:dedisp}).  If scattering is significant in the CHIME range, CHIME/FRB will see clear evolution of the pulse morphology over the band, which spans a factor of $2^{4.4} \simeq 21$ in broadening time.  The detection of significant scattering — beyond that expected from the Milky Way alone — in a large fraction of FRBs would argue strongly that the population is found in disks of spiral galaxies, or special regions of high plasma turbulence such as near supermassive black holes, or in supernova remnants.

Another possible reason for a reduced FRB rate in the CHIME band is free-free absorption.   This effect is important in dense, ionized plasma where some models, notably those involving young neutron stars in supernova remnants \cite[e.g.][]{piro16}, have proposed FRBs are located. The free-free optical depth varies as $\tau_{ff} \propto {\rm EM}\,\nu^{-2.1}$, where EM is the emission measure $\int n_e^2 dl$, with $n_e$ the electron density.  Since the absorbed radio flux has intensity that varies as $e^{-\tau_{ff}}$, even a relatively small $\tau_{ff}$ can lead to fading toward the lower end of the spectrum, hence a steep inverted spectrum \citep[see][for a discussion]{kon+14}.  This would manifest itself as a rapid fading in the CHIME band, with width otherwise constant, in contrast to pure scattering for which fluence remains constant but pulse width increases. However, there are no hints of absorption in the vast majority of FRB bursts at 1.4 GHz.  Though FRB 121102 has shown steep inverted spectra, these have been established as being intrinsic to the source, whose spectrum is highly variable, with steeply rising spectra observed as commonly as falling ones \citep{ssh+16a,ssh+16b}.
The detection of significant free-free absorption in many FRB spectra would indicate that the population inhabits regions of high plasma densities, strongly suggesting a very young population. 

Finally, on average, FRB spectra may be intrinsically fainter at lower radio frequencies. Without a detailed source emission model, the likelihood of this cannot be determined. This possibility can be distinguished from the free-free absorption case by CHIME/FRB observations, since the latter should follow strictly the predicted free-free frequency dependence, and may also show a correlation with DM after correcting for the Milky Way and intergalactic contributions. There is no reason to think an intrinsic spectral change should have the same frequency dependence as free-free absorption, or be correlated with source environment. 

\citet{ckj+17} have done simulations of FRB populations subject to various assumptions regarding intrinsic spectral index, free-free absorption, and scattering measure distributions. For example, they have predicted CHIME/FRB event rates of order 10 per day for a Euclidean flux distribution and a scattering-measure distribution like that observed in the Milky Way, with a limiting intrinsic spectral index constrained by the absence of detections at 350 MHz in the GBNCC survey \citep[see][for details]{ckj+17}.  The uncertainty is an order of magnitude  in the 700--800-MHz band, and approximately two orders in the 400--500-MHz band. The predicted CHIME/FRB event rates from \citet{ckj+17} are broadly consistent with those based on a simpler analysis by \citet{clm+16} who considered the 700--900 MHz band only.  

\begin{figure}[t]
	\includegraphics[width=\textwidth]{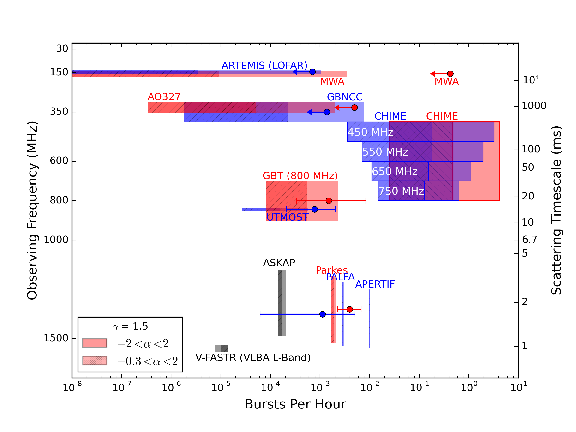}
    \caption{Estimated CHIME/FRB event detection rate, displayed alongside those of other FRB detection efforts, with radio frequency on the vertical axis.  This represents an update from a similar plot presented by \citet{ckj+17} but based on the revised all-sky rate estimated by \citet{lvl+17}.  The red CHIME box is for the full CHIME bandwidth, whereas the blue rectangles show estimates in four 100-MHz sub-bands. A Euclidean flux distribution ($\gamma=1.5$) is assumed. The hatched boxes denote rate estimates assuming a spectral index range, $-0.3 < \alpha < 2$, that was estimated in \citet{ckj+17}.  Different colors are used to distinguish different experiments in the same frequency range. }
    \label{fig:rate}
\end{figure}

However, the predictions presented in \citet{ckj+17} are derived from the \citet{crt+16} 1.4-GHz FRB rate which is a factor of several higher than the more recent estimate by \citet{lvl+17} of 587$^{+924}_{-272}$ bursts sky$^{-1}$ day$^{-1}$ above a peak flux density of 1 Jy. Based on this revised rate and assuming a Euclidean flux distribution, our revised analysis predicts CHIME/FRB will detect 0.6--11 bursts per day, as shown in Figure~\ref{fig:rate}. This prediction is obtained using simulations following the same methodology as in \citet{ckj+17} except that the simulated bursts have an intrinsic width of 3 ms with scattering timescales drawn from a distribution similar to the Galactic scattering-time distribution at 1 GHz. A significant reduction in the detection rate (0.3--2 bursts per day) is expected if FRBs are detectable only in the upper part of the CHIME band (700--800 MHz).  Although these updated CHIME/FRB rate predictions are significantly lower than previous estimates, it is important to note that \fix{our} revised rates have been calculated using an all-sky event rate. Actual detection rates could be as much as a factor of four higher since \fix{75\%} of the CHIME-visible sky corresponds to high Galactic latitudes ($|b| > 15^{\circ}$) where the observed rate is estimated to be an order of magnitude greater than the low-latitude rate: 2866$^{+7328}_{-1121}$ bursts sky$^{-1}$ day$^{-1}$ for a threshold of 1 Jy \citep{vbl+16}.  This implies a CHIME/FRB rate of 2--42 per day.

If CHIME/FRB detects many FRBs per day, thousands of FRBs over the currently envisioned $\sim$3~yr lifetime of the project (presently limited only by available operations funding) will be detected.  This will allow detailed sky distributions, DM distributions and a log$N$/log$S$ distribution with the benefits of a uniform survey with a single well determined sensitivity function. This will also enable the study of potential correlations between parameters such as DM and flux, DM and scattering measure, or width and flux. The FRB DM distribution, for example, has been suggested to be informative of the location of the so-called missing cosmic baryons \citep{mcq14} and could provide a new probe of the  three dimensional clustering of matter in the universe \citep{ms15}. \citet{yz16} suggest that, given a large number of FRBs with redshift determinations, one can infer cosmological parameters even without knowing the host galaxy DM contribution.  The detection of many FRBs at very high redshifts ($z > 6$) could allow a determination of the size distribution of ionized bubbles in the intergalactic medium near the Epoch of Reionization \citep{yt18}. 
Sky distributions of FRBs will also allow us to understand Galactic propagation effects such as plasma lensing \citep{cwh+17} and separate those from intrinsic FRB characteristics.
Additionally, we can determine robust upper limits on short radio transient emission in the CHIME FoV in the absence of a detection by our pipeline.  This will be useful for setting upper limits on radio emission from counterparts from sources that trigger e.g. gravitational wave detectors or other electromagnetic transient detectors.

On the other hand, CHIME/FRB's angular resolution alone is insufficient for host-galaxy identification, a key ingredient in making progress in this field.  However, there are several potential paths to improved localization and/or redshift determination. Because CHIME sees the full sky above declination $\sim -20^{\circ}$ each day, CHIME/FRB will be sensitive to repeat bursts, enabling identification of repeating FRBs like FRB 121102.  These can be followed up by long baseline interferometers for localization and subsequent host galaxy identification as was done for FRB 121102 \citep{clw+17,tbc+17}.  Note that statistics of repeaters will be well studied thanks to CHIME/FRB's regular daily exposure of fixed duration with stable sensitivity. Another avenue for localization will be through VLBI using the CHIME/FRB baseband recording, in conjunction with another telescope operating in the CHIME band and watching the same portion of sky.  ``Outrigger'' stations are presently under study.

One way CHIME/FRB could potentially determine a host galaxy redshift independently is through H\textsc{i} absorption of a bright FRB's spectrum, from the host galaxy's hydrogen gas. For an FRB in the redshift range 0.8--2.5, assuming a flux of 30~Jy, the absorption is expected to have SNR $\gapp$ 15  approximately 10\% of the time, assuming FRBs are in gas-rich spiral galaxies viewed randomly \citep{ml16}. Note that FRBs with fluxes of up to 120 Jy (albeit at 1.4 GHz) have been reported \citep[e.g.][]{rsb+16}. The above SNR estimate assumes the planned 16k frequency channels across the CHIME bandwidth, but with baseband data we could improve the resolution further, rendering detection easier. Moreover, for repeating FRBs, we can stack spectra to improve the SNR of a possible absorption feature. However, a redshift detection via H\textsc{i} absorption requires that not all FRBs be in dwarf galaxies, \fix{as is the case for FRB 121102 \citep{tbc+17},} for which the effect would likely be undetectable.

%
%

\section{Conclusions}

The CHIME/FRB project has the potential to yield major progress in our understanding of the FRB mystery. This is remarkable for a custom-designed telescope whose basic properties were chosen explicitly for independent science. The telescope's utility for simultaneous pulsar timing via CHIME/Pulsar is further evidence of the scientific value of a ``software" radio telescope given today's massive computing capabilities.  Computing power and storage are likely to grow given massive consumer demand, suggesting CHIME is easily scalable, which, in the FRB context, argues that far deeper surveys may one day be possible. Whether efforts in this direction are well motivated will be determined by FRBs' utility as cosmic probes, something CHIME/FRB should be able to resolve.  Crucial to this endeavour is the ability to follow-up FRBs to identify host galaxies and redshifts, something CHIME/FRB cannot do alone because of its relatively poor localization capabilities.  However, interferometric follow-up, especially of repeating FRBs, which CHIME/FRB is well placed to discover, is a promising avenue for localization (as has been accomplished for FRB 121102) and is being pursued by our team.  Another attractive option, especially for non-repeaters, is the correlation of CHIME/FRB baseband data with those obtained from an array of CHIME/FRB ``outriggers'' located across North America. Such an effort is presently under consideration.  Another future avenue for enhancement of CHIME/FRB's baseline capabilities is also under investigation:  the use of machine-learning algorithms at various places in the CHIME/FRB pipeline, namely for improved RFI mitigation, improved real-time as well as offline event identification. This may prove to be yet another manifestation of the power of data-rich software telescopes in that they permit great creativity and flexibility in data handling and hence in scientific output.

\bigskip

We are very grateful for the warm reception and skillful help
we have received from the staff of the Dominion Radio
Astrophysical Observatory, which is operated by the National
Research Council Canada.
The CHIME/FRB Project is funded by a grant from the Canada Foundation for Innovation
2015 Innovation Fund (Project 33213), as well as by the Provinces of British Columbia
and Quebec.  Additional support was provided by the Canadian Institute for Advanced Research (CIFAR) Gravity \& Extreme Universe Program, McGill University and the McGill Space Institute, University of British Columbia, and
University of Toronto Dunlap Institute.
The Dunlap Institute is funded by an endowment established by the David Dunlap family and the University of Toronto.
Research at Perimeter Institute is supported by the Government of Canada through Industry Canada and by the Province of Ontario through the Ministry of Research \& Innovation.
The National Radio Astronomy Observatory is a facility of the National Science Foundation operated under cooperative agreement by Associated Universities, Inc.
J.C. was supported by a Natural Sciences and Engineering Research Council of Canada (NSERC) Undergraduate Student Research Award.
P.C. is supported by a Mitacs Globalink Graduate Fellowship.
M.D. is supported by an NSERC Discovery Grant, CIFAR and Fonds de Recherche du Qu\'ebec -- Nature et technologies Centre de Recherche en Astrophysique du Qu\'ebec (FRQNT/CRAQ).
B.M.G. is supported by an NSERC Discovery Grant and the Canada Research Chairs program.
M.H. and G.H. are supported by CIFAR.
V.M.K. is supported by a Lorne Trottier Chair in Astrophysics \& Cosmology, a Canada Research Chair, CIFAR, by an NSERC Discovery Grant and Herzberg Award, and by FRQNT/CRAQ.
U.-L.P. is supported by CIFAR.
K.W.M. is supported by the Canadian Institute for Theoretical Astrophysics National Fellows program.
S.M.R. is supported by CIFAR and the NSF Physics Frontiers Center award 1430284.
I.H.S. is supported by CIFAR and an NSERC Discovery Grant.
P.S. is supported by a DRAO Covington Fellowship from the National Research Council Canada.

\bibliography{journals1,modrefs,crossrefs,frbrefs,psrrefs}

\begin{thebibliography}{}
\expandafter\ifx\csname natexlab\endcsname\relax\def\natexlab#1{#1}\fi
\providecommand{\url}[1]{\href{#1}{#1}}

\bibitem[{{Akahori} {et~al.}(2016){Akahori}, {Ryu}, \& {Gaensler}}]{arg16}
{Akahori}, T., {Ryu}, D., \& {Gaensler}, B.~M. 2016, \apj, 824, 105

\bibitem[{{Amiri} {et~al.}(2017){Amiri}, {Bandura}, {Berger}, {Bond}, {Cliche},
  {Connor}, {Deng}, {Denman}, {Dobbs}, {Domagalski}, {Fandino}, {Gilbert},
  {Good}, {Halpern}, {Hanna}, {Hincks}, {Hinshaw}, {H{\"o}fer}, {Hsyu},
  {Klages}, {Landecker}, {Masui}, {Mena-Parra}, {Newburgh}, {Oppermann}, {Pen},
  {Peterson}, {Pinsonneault-Marotte}, {Renard}, {Shaw}, {Siegel}, {Sigurdson},
  {Smith}, {Storer}, {Tretyakov}, {Vanderlinde}, {Wiebe}, \& {Scientific
  Collaboration20}}]{abb+17}
{Amiri}, M., {Bandura}, K., {Berger}, P., {et~al.} 2017, \apj, 844, 161

\bibitem[{{Bandura} {et~al.}(2016{\natexlab{a}}){Bandura}, {Cliche}, {Dobbs},
  {Gilbert}, {Ittah}, {Mena Parra}, \& {Smecher}}]{bcd+16}
{Bandura}, K., {Cliche}, J.~F., {Dobbs}, M.~A., {et~al.} 2016{\natexlab{a}},
  Journal of Astronomical Instrumentation, 5, 1641004

\bibitem[{{Bandura} {et~al.}(2014){Bandura}, {Addison}, {Amiri}, {Bond},
  {Campbell-Wilson}, {Connor}, {Cliche}, {Davis}, {Deng}, {Denman}, {Dobbs},
  {Fandino}, {Gibbs}, {Gilbert}, {Halpern}, {Hanna}, {Hincks}, {Hinshaw},
  {H{\"o}fer}, {Klages}, {Landecker}, {Masui}, {Mena Parra}, {Newburgh}, {Pen},
  {Peterson}, {Recnik}, {Shaw}, {Sigurdson}, {Sitwell}, {Smecher}, {Smegal},
  {Vanderlinde}, \& {Wiebe}}]{baa+15}
{Bandura}, K., {Addison}, G.~E., {Amiri}, M., {et~al.} 2014, Proc. SPIE, 9145,
  914522

\bibitem[{{Bandura} {et~al.}(2016{\natexlab{b}}){Bandura}, {Bender}, {Cliche},
  {de Haan}, {Dobbs}, {Gilbert}, {Griffin}, {Hsyu}, {Ittah}, {Parra},
  {Montgomery}, {Pinsonneault-Marotte}, {Siegel}, {Smecher}, {Tang},
  {Vanderlinde}, \& {Whitehorn}}]{bbc+16}
{Bandura}, K., {Bender}, A.~N., {Cliche}, J.~F., {et~al.} 2016{\natexlab{b}},
  Journal of Astronomical Instrumentation, 5, 1641005

\bibitem[{{Bannister} {et~al.}(2017){Bannister}, {Shannon}, {Macquart},
  {Flynn}, {Edwards}, {O'Neill}, {Os{\l}owski}, {Bailes}, {Zackay}, {Clarke},
  {D'Addario}, {Dodson}, {Hall}, {Jameson}, {Jones}, {Navarro}, {Trinh},
  {Allison}, {Anderson}, {Bell}, {Chippendale}, {Collier}, {Heald}, {Heywood},
  {Hotan}, {Lee-Waddell}, {Madrid}, {Marvil}, {McConnell}, {Popping},
  {Voronkov}, {Whiting}, {Allen}, {Bock}, {Brodrick}, {Cooray}, {DeBoer},
  {Diamond}, {Ekers}, {Gough}, {Hampson}, {Harvey-Smith}, {Hay}, {Hayman},
  {Jackson}, {Johnston}, {Koribalski}, {McClure-Griffiths}, {Mirtschin}, {Ng},
  {Norris}, {Pearce}, {Phillips}, {Roxby}, {Troup}, \& {Westmeier}}]{bsm+17}
{Bannister}, K.~W., {Shannon}, R.~M., {Macquart}, J.-P., {et~al.} 2017, \apjl,
  841, L12

\bibitem[{{Caleb} {et~al.}(2017){Caleb}, {Flynn}, {Bailes}, {Barr}, {Bateman},
  {Bhandari}, {Campbell-Wilson}, {Farah}, {Green}, {Hunstead}, {Jameson},
  {Jankowski}, {Keane}, {Parthasarathy}, {Ravi}, {Rosado}, {van Straten}, \&
  {Venkatraman Krishnan}}]{cfb+17}
{Caleb}, M., {Flynn}, C., {Bailes}, M., {et~al.} 2017, \mnras, 468, 3746

\bibitem[{{Chatterjee} {et~al.}(2017){Chatterjee}, {Law}, {Wharton},
  {Burke-Spolaor}, {Hessels}, {Bower}, {Cordes}, {Tendulkar}, {Bassa},
  {Demorest}, {Butler}, {Seymour}, {Scholz}, {Abruzzo}, {Bogdanov}, {Kaspi},
  {Keimpema}, {Lazio}, {Marcote}, {McLaughlin}, {Paragi}, {Ransom}, {Rupen},
  {Spitler}, \& {van Langevelde}}]{clw+17}
{Chatterjee}, S., {Law}, C.~J., {Wharton}, R.~S., {et~al.} 2017, \nat, 541, 58

\bibitem[{{Chawla} {et~al.}(2017){Chawla}, {Kaspi}, {Josephy}, {Rajwade},
  {Lorimer}, {Archibald}, {DeCesar}, {Hessels}, {Kaplan}, {Karako-Argaman},
  {Kondratiev}, {Levin}, {Lynch}, {McLaughlin}, {Ransom}, {Roberts}, {Stairs},
  {Stovall}, {Swiggum}, \& {van Leeuwen}}]{ckj+17}
{Chawla}, P., {Kaspi}, V.~M., {Josephy}, A., {et~al.} 2017, \apj, 844, 140

\bibitem[{{Connor} {et~al.}(2016){Connor}, {Lin}, {Masui}, {Oppermann}, {Pen},
  {Peterson}, {Roman}, \& {Sievers}}]{clm+16}
{Connor}, L., {Lin}, H.-H., {Masui}, K., {et~al.} 2016, \mnras, 460, 1054

\bibitem[{{Cordes} \& {Lazio}(2001)}]{cl01}
{Cordes}, J.~M., \& {Lazio}, T.~J.~W. 2001, ApJ, 549, 997

\bibitem[{Cordes \& {McLaughlin}(2003)}]{cm03}
Cordes, J.~M., \& {McLaughlin}, M.~A. 2003, ApJ, 596, 1142

\bibitem[{{Cordes} {et~al.}(2017){Cordes}, {Wasserman}, {Hessels}, {Lazio},
  {Chatterjee}, \& {Wharton}}]{cwh+17}
{Cordes}, J.~M., {Wasserman}, I., {Hessels}, J.~W.~T., {et~al.} 2017, \apj,
  842, 35

\bibitem[{Cortes \& Vapnik(1995)}]{cv95a}
Cortes, C., \& Vapnik, V. 1995, Machine Learning, 20, 273

\bibitem[{{Crawford} {et~al.}(2016){Crawford}, {Rane}, {Tran}, {Rolph},
  {Lorimer}, \& {Ridley}}]{crt+16}
{Crawford}, F., {Rane}, A., {Tran}, L., {et~al.} 2016, \mnras, 460, 3370

\bibitem[{{Deng} {et~al.}(2014){Deng}, {Campbell-Wilson}, \& {CHIME
  Collaboration}}]{mdc+14}
{Deng}, M., {Campbell-Wilson}, D., \& {CHIME Collaboration}, T. 2014, ``2014
  16th International Symposium on Antenna Technology and Applied
  Electromagnetics (ANTEM)", 1

\bibitem[{{Eatough} {et~al.}(2009){Eatough}, {Keane}, \& {Lyne}}]{ekl09}
{Eatough}, R.~P., {Keane}, E.~F., \& {Lyne}, A.~G. 2009, \mnras, 395, 410

\bibitem[{Ester {et~al.}(1996)Ester, Kriegal, Sander, Xu, Simoudis, Han, \&
  Fayyad}]{eks+96}
Ester, M., Kriegal, H.-P., Sander, J., {et~al.} 1996, A Density-based Algorithm
  for Discovering Clusters in Large Spatial Databases with Noise

\bibitem[{{Fialkov} \& {Loeb}(2016)}]{fl16}
{Fialkov}, A., \& {Loeb}, A. 2016, \apj, 821, 59

\bibitem[{{Hankins} \& {Rickett}(1975)}]{hr75}
{Hankins}, T.~H., \& {Rickett}, B.~J. 1975, in Methods in Computational Physics
  Volume 14 --- Radio Astronomy (New York: Academic Press), 55--129

\bibitem[{{Hincks} {et~al.}(2015){Hincks}, {Shaw}, \& {CHIME
  Collaboration}}]{hsc+14}
{Hincks}, A.~D., {Shaw}, J.~R., \& {CHIME Collaboration}. 2015, in Astronomical
  Society of the Pacific Conference Series, Vol. 495, Astronomical Data
  Analysis Software an Systems XXIV (ADASS XXIV), ed. A.~R. {Taylor} \&
  E.~{Rosolowsky}, 523

\bibitem[{{Inoue}(2004)}]{ino04}
{Inoue}, S. 2004, \mnras, 348, 999

\bibitem[{{Karako-Argaman} {et~al.}(2015){Karako-Argaman}, {Kaspi}, {Lynch},
  {Hessels}, {Kondratiev}, {McLaughlin}, {Ransom}, {Archibald}, {Boyles},
  {Jenet}, {Kaplan}, {Levin}, {Lorimer}, {Madsen}, {Roberts}, {Siemens},
  {Stairs}, {Stovall}, {Swiggum}, \& {van Leeuwen}}]{kkl+15}
{Karako-Argaman}, C., {Kaspi}, V.~M., {Lynch}, R.~S., {et~al.} 2015, \apj, 809,
  67

\bibitem[{{Karastergiou} {et~al.}(2015){Karastergiou}, {Chennamangalam},
  {Armour}, {Williams}, {Mort}, {Dulwich}, {Salvini}, {Magro}, {Roberts},
  {Serylak}, {Doo}, {Bilous}, {Breton}, {Falcke}, {Grie{\ss}meier}, {Hessels},
  {Keane}, {Kondratiev}, {Kramer}, {van Leeuwen}, {Noutsos}, {Os{\l}owski},
  {Sobey}, {Stappers}, \& {Weltevrede}}]{kca+15}
{Karastergiou}, A., {Chennamangalam}, J., {Armour}, W., {et~al.} 2015, \mnras,
  452, 1254

\bibitem[{{Kothes} {et~al.}(2010){Kothes}, {Landecker}, \& {Gray}}]{klg10}
{Kothes}, R., {Landecker}, T.~L., \& {Gray}, A.~D. 2010, in Astronomical
  Society of the Pacific Conference Series, Vol. 438, The Dynamic Interstellar
  Medium: A Celebration of the Canadian Galactic Plane Survey, ed. R.~{Kothes},
  T.~L. {Landecker}, \& A.~G. {Willis}, 415

\bibitem[{{Krishnakumar} {et~al.}(2015){Krishnakumar}, {Mitra}, {Naidu},
  {Joshi}, \& {Manoharan}}]{kmn+15}
{Krishnakumar}, M.~A., {Mitra}, D., {Naidu}, A., {Joshi}, B.~C., \&
  {Manoharan}, P.~K. 2015, \apj, 804, 23

\bibitem[{{Kulkarni} {et~al.}(2014){Kulkarni}, {Ofek}, {Neill}, {Zheng}, \&
  {Juric}}]{kon+14}
{Kulkarni}, S.~R., {Ofek}, E.~O., {Neill}, J.~D., {Zheng}, Z., \& {Juric}, M.
  2014, \apj, 797, 70

\bibitem[{{Lawrence} {et~al.}(2017){Lawrence}, {Vander Wiel}, {Law}, {Burke
  Spolaor}, \& {Bower}}]{lvl+17}
{Lawrence}, E., {Vander Wiel}, S., {Law}, C., {Burke Spolaor}, S., \& {Bower},
  G.~C. 2017, \aj, 154, 117

\bibitem[{{Lorimer} {et~al.}(2007){Lorimer}, {Bailes}, {McLaughlin},
  {Narkevic}, \& {Crawford}}]{lbm+07}
{Lorimer}, D.~R., {Bailes}, M., {McLaughlin}, M.~A., {Narkevic}, D.~J., \&
  {Crawford}, F. 2007, Science, 318, 777

\bibitem[{{Manchester} {et~al.}(2005){Manchester}, {Hobbs}, {Teoh}, \&
  {Hobbs}}]{mhth05}
{Manchester}, R.~N., {Hobbs}, G.~B., {Teoh}, A., \& {Hobbs}, M. 2005, AJ, 129,
  1993

\bibitem[{Manchester \& Taylor(1977)}]{mt77}
Manchester, R.~N., \& Taylor, J.~H. 1977, Pulsars (San Francisco: Freeman)

\bibitem[{{Maoz} {et~al.}(2015){Maoz}, {Loeb}, {Shvartzvald}, {Sitek}, {Engel},
  {Kiefer}, {Kiraga}, {Levi}, {Mazeh}, {Pawlak}, {Rich}, {Tal-Or}, \&
  {Wyrzykowski}}]{mls+15b}
{Maoz}, D., {Loeb}, A., {Shvartzvald}, Y., {et~al.} 2015, \mnras, 454, 2183

\bibitem[{{Marcote} {et~al.}(2017){Marcote}, {Paragi}, {Hessels}, {Keimpema},
  {van Langevelde}, {Huang}, {Bassa}, {Bogdanov}, {Bower}, {Burke-Spolaor},
  {Butler}, {Campbell}, {Chatterjee}, {Cordes}, {Demorest}, {Garrett}, {Ghosh},
  {Kaspi}, {Law}, {Lazio}, {McLaughlin}, {Ransom}, {Salter}, {Scholz},
  {Seymour}, {Siemion}, {Spitler}, {Tendulkar}, \& {Wharton}}]{mph+17}
{Marcote}, B., {Paragi}, Z., {Hessels}, J.~W.~T., {et~al.} 2017, \apjl, 834, L8

\bibitem[{{Margalit} \& {Loeb}(2016)}]{ml16}
{Margalit}, B., \& {Loeb}, A. 2016, \mnras, 460, L25

\bibitem[{{Masui} {et~al.}(2015){Masui}, {Lin}, {Sievers}, {Anderson}, {Chang},
  {Chen}, {Ganguly}, {Jarvis}, {Kuo}, {Li}, {Liao}, {McLaughlin}, {Pen},
  {Peterson}, {Roman}, {Timbie}, {Voytek}, \& {Yadav}}]{mls+15a}
{Masui}, K., {Lin}, H.-H., {Sievers}, J., {et~al.} 2015, \nat, 528, 523

\bibitem[{{Masui} {et~al.}(2017){Masui}, {Shaw}, {Ng}, {Smith}, {Vanderlinde},
  \& {Paradise}}]{msn+17}
{Masui}, K.~W., {Shaw}, J.~R., {Ng}, C., {et~al.} 2017, ArXiv e-prints,
  arXiv:1710.08591

\bibitem[{{Masui} \& {Sigurdson}(2015)}]{ms15}
{Masui}, K.~W., \& {Sigurdson}, K. 2015, Physical Review Letters, 115, 121301

\bibitem[{{McLaughlin} \& {Cordes}(2003)}]{mc03}
{McLaughlin}, M.~A., \& {Cordes}, J.~M. 2003, ApJ, 596, 982

\bibitem[{{McLaughlin} {et~al.}(2006){McLaughlin}, {Lyne}, {Lorimer}, {Kramer},
  {Faulkner}, {Manchester}, {Cordes}, {Camilo}, {Possenti}, {Stairs}, {Hobbs},
  {D'Amico}, {Burgay}, \& {O'Brien}}]{mll+06}
{McLaughlin}, M.~A., {Lyne}, A.~G., {Lorimer}, D.~R., {et~al.} 2006, Nature,
  439, 817

\bibitem[{McQuinn(2014)}]{mcq14}
McQuinn, M. 2014, \apj, 780, L33

\bibitem[{{Michilli} {et~al.}(2018){Michilli}, {Seymour}, {Hessels}, {Spitler},
  {Gajjar}, {Archibald}, {Bower}, {Chatterjee}, {Cordes}, {Gourdji}, {Heald},
  {Kaspi}, {Law}, {Sobey}, {Adams}, {Bassa}, {Bogdanov}, {Brinkman},
  {Demorest}, {Fernandez}, {Hellbourg}, {Lazio}, {Lynch}, {Maddox}, {Marcote},
  {McLaughlin}, {Paragi}, {Ransom}, {Scholz}, {Siemion}, {Tendulkar}, {Van
  Rooy}, {Wharton}, \& {Whitlow}}]{msh+18}
{Michilli}, D., {Seymour}, A., {Hessels}, J.~W.~T., {et~al.} 2018, Nature, 533,
  132

\bibitem[{{Newburgh} {et~al.}(2014){Newburgh}, {Addison}, {Amiri}, {Bandura},
  {Bond}, {Connor}, {Cliche}, {Davis}, {Deng}, {Denman}, {Dobbs}, {Fandino},
  {Fong}, {Gibbs}, {Gilbert}, {Griffin}, {Halpern}, {Hanna}, {Hincks},
  {Hinshaw}, {H{\"o}fer}, {Klages}, {Landecker}, {Masui}, {Parra}, {Pen},
  {Peterson}, {Recnik}, {Shaw}, {Sigurdson}, {Sitwell}, {Smecher}, {Smegal},
  {Vanderlinde}, \& {Wiebe}}]{naa+14}
{Newburgh}, L.~B., {Addison}, G.~E., {Amiri}, M., {et~al.} 2014, Proc. SPIE,
  9145, 91454V

\bibitem[{{Ng} \& {CHIME/Pulsar~Collaboration}(2017)}]{n++17}
{Ng}, C., \& {CHIME/Pulsar~Collaboration}. 2017, in Pulsar Astrophysics -- The
  Next 50 Years, Proceedings of IAU Symposium No. 337, in press

\bibitem[{{Ng} {et~al.}(2017){Ng}, {Vanderlinde}, {Paradise}, {Klages},
  {Masui}, {Smith}, {Bandura}, {Boyle}, {Dobbs}, {Kaspi}, {Renard}, {Shaw},
  {Stairs}, \& {Tretyakov}}]{nvp+17}
{Ng}, C., {Vanderlinde}, K., {Paradise}, A., {et~al.} 2017, ArXiv e-prints,
  arXiv:1702.04728

\bibitem[{{Opperman} \& {Pen}(2017)}]{op17}
{Opperman}, N., \& {Pen}, U.-L. 2017, ArXiv e-prints, arXiv:1705.04881

\bibitem[{Pedregosa {et~al.}(2011)Pedregosa, Varoquaux, Gramfort, Michel,
  Thirion, Grisel, Blondel, Prettenhofer, Weiss, Dubourg, Vanderplas, Passos,
  Cournapeau, Brucher, Perrot, \& Duchesnay}]{pvg+11}
Pedregosa, F., Varoquaux, G., Gramfort, A., {et~al.} 2011, Journal of Machine
  Learning Research, 12, 2825

\bibitem[{{Petroff} {et~al.}(2015{\natexlab{a}}){Petroff}, {Keane}, {Barr},
  {Reynolds}, {Sarkissian}, {Edwards}, {Stevens}, {Brem}, {Jameson},
  {Burke-Spolaor}, {Johnston}, {Bhat}, {Kudale}, \& {Bhandari}}]{pkb+15}
{Petroff}, E., {Keane}, E.~F., {Barr}, E.~D., {et~al.} 2015{\natexlab{a}},
  \mnras, 451, 3933

\bibitem[{{Petroff} {et~al.}(2015{\natexlab{b}}){Petroff}, {Johnston}, {Keane},
  {van Straten}, {Bailes}, {Barr}, {Barsdell}, {Burke-Spolaor}, {Caleb},
  {Champion}, {Flynn}, {Jameson}, {Kramer}, {Ng}, {Possenti}, \&
  {Stappers}}]{pjk+15}
{Petroff}, E., {Johnston}, S., {Keane}, E.~F., {et~al.} 2015{\natexlab{b}},
  \mnras, 454, 457

\bibitem[{{Petroff} {et~al.}(2015{\natexlab{c}}){Petroff}, {Bailes}, {Barr},
  {Barsdell}, {Bhat}, {Bian}, {Burke-Spolaor}, {Caleb}, {Champion}, {Chandra},
  {Da Costa}, {Delvaux}, {Flynn}, {Gehrels}, {Greiner}, {Jameson}, {Johnston},
  {Kasliwal}, {Keane}, {Keller}, {Kocz}, {Kramer}, {Leloudas}, {Malesani},
  {Mulchaey}, {Ng}, {Ofek}, {Perley}, {Possenti}, {Schmidt}, {Shen},
  {Stappers}, {Tisserand}, {van Straten}, \& {Wolf}}]{pbb+15}
{Petroff}, E., {Bailes}, M., {Barr}, E.~D., {et~al.} 2015{\natexlab{c}},
  \mnras, 447, 246

\bibitem[{{Petroff} {et~al.}(2016){Petroff}, {Barr}, {Jameson}, {Keane},
  {Bailes}, {Kramer}, {Morello}, {Tabbara}, \& {van Straten}}]{pbj+16}
{Petroff}, E., {Barr}, E.~D., {Jameson}, A., {et~al.} 2016, \pasa, 33, e045

\bibitem[{{Petroff} {et~al.}(2017){Petroff}, {Houben}, {Bannister},
  {Burke-Spolaor}, {Cordes}, {Falcke}, {van Haren}, {Karastergiou}, {Kramer},
  {Law}, {van Leeuwen}, {Lorimer}, {Martinez-Rubi}, {Rachen}, {Spitler}, \&
  {Weltman}}]{phb+17}
{Petroff}, E., {Houben}, L., {Bannister}, K., {et~al.} 2017, ArXiv e-prints,
  arXiv:1710.08155

\bibitem[{{Piro}(2016)}]{piro16}
{Piro}, A.~L. 2016, \apjl, 824, L32

\bibitem[{{Prochaska} \& {Neeleman}(2018)}]{pn18}
{Prochaska}, J.~X., \& {Neeleman}, M. 2018, \mnras, 474, 318

\bibitem[{{Rane} \& {Lorimer}(2017)}]{rl17}
{Rane}, A., \& {Lorimer}, D. 2017, Journal of Astrophysics and Astronomy, 38,
  55

\bibitem[{{Rane} {et~al.}(2016){Rane}, {Lorimer}, {Bates}, {McMann},
  {McLaughlin}, \& {Rajwade}}]{rlb+16}
{Rane}, A., {Lorimer}, D.~R., {Bates}, S.~D., {et~al.} 2016, \mnras, 455, 2207

\bibitem[{{Ravi} {et~al.}(2016){Ravi}, {Shannon}, {Bailes}, {Bannister},
  {Bhandari}, {Bhat}, {Burke-Spolaor}, {Caleb}, {Flynn}, {Jameson}, {Johnston},
  {Keane}, {Kerr}, {Tiburzi}, {Tuntsov}, \& {Vedantham}}]{rsb+16}
{Ravi}, V., {Shannon}, R.~M., {Bailes}, M., {et~al.} 2016, Science, 354, 1249

\bibitem[{{Remazeilles} {et~al.}(2015){Remazeilles}, {Dickinson}, {Banday},
  {Bigot-Sazy}, \& {Ghosh}}]{rdb+15}
{Remazeilles}, M., {Dickinson}, C., {Banday}, A.~J., {Bigot-Sazy}, M.-A., \&
  {Ghosh}, T. 2015, \mnras, 451, 4311

\bibitem[{Rickett(1977)}]{ric77}
Rickett, B.~J. 1977, ARAA, 15, 479

\bibitem[{{Rowlinson} {et~al.}(2016){Rowlinson}, {Bell}, {Murphy}, {Trott},
  {Hurley-Walker}, {Johnston}, {Tingay}, {Kaplan}, {Carbone}, {Hancock},
  {Feng}, {Offringa}, {Bernardi}, {Bowman}, {Briggs}, {Cappallo}, {Deshpande},
  {Gaensler}, {Greenhill}, {Hazelton}, {Johnston-Hollitt}, {Lonsdale},
  {McWhirter}, {Mitchell}, {Morales}, {Morgan}, {Oberoi}, {Ord}, {Prabu},
  {Udaya Shankar}, {Srivani}, {Subrahmanyan}, {Wayth}, {Webster}, {Williams},
  \& {Williams}}]{rbm+16}
{Rowlinson}, A., {Bell}, M.~E., {Murphy}, T., {et~al.} 2016, \mnras, 458, 3506

\bibitem[{{Scholz} {et~al.}(2016){Scholz}, {Spitler}, {Hessels}, {Chatterjee},
  {Cordes}, {Kaspi}, {Wharton}, {Bassa}, {Bogdanov}, {Camilo}, {Crawford},
  {Deneva}, {van Leeuwen}, {Lynch}, {Madsen}, {McLaughlin}, {Mickaliger},
  {Parent}, {Patel}, {Ransom}, {Seymour}, {Stairs}, {Stappers}, \&
  {Tendulkar}}]{ssh+16b}
{Scholz}, P., {Spitler}, L.~G., {Hessels}, J.~W.~T., {et~al.} 2016, \apj, 833,
  177

\bibitem[{{Shull} \& {Danforth}(2018)}]{sd18}
{Shull}, J.~M., \& {Danforth}, C.~W. 2018, \apjl, 852, L11

\bibitem[{Smith(et al. in prep.)}]{s++18}
Smith, K.~M. et al. in prep.

\bibitem[{{Spitler} {et~al.}(2014){Spitler}, {Cordes}, {Hessels}, {Lorimer},
  {McLaughlin}, {Chatterjee}, {Crawford}, {Deneva}, {Kaspi}, {Wharton},
  {Allen}, {Bogdanov}, {Brazier}, {Camilo}, {Freire}, {Jenet},
  {Karako-Argaman}, {Knispel}, {Lazarus}, {Lee}, {van Leeuwen}, {Lynch},
  {Ransom}, {Scholz}, {Siemens}, {Stairs}, {Stovall}, {Swiggum},
  {Venkataraman}, {Zhu}, {Aulbert}, \& {Fehrmann}}]{sch+14}
{Spitler}, L.~G., {Cordes}, J.~M., {Hessels}, J.~W.~T., {et~al.} 2014, \apj,
  790, 101

\bibitem[{{Spitler} {et~al.}(2016){Spitler}, {Scholz}, {Hessels}, {Bogdanov},
  {Brazier}, {Camilo}, {Chatterjee}, {Cordes}, {Crawford}, {Deneva}, {Ferdman},
  {Freire}, {Kaspi}, {Lazarus}, {Lynch}, {Madsen}, {McLaughlin}, {Patel},
  {Ransom}, {Seymour}, {Stairs}, {Stappers}, {van Leeuwen}, \& {Zhu}}]{ssh+16a}
{Spitler}, L.~G., {Scholz}, P., {Hessels}, J.~W.~T., {et~al.} 2016, \nat, 531,
  202

\bibitem[{{Taylor}(1974)}]{tay74}
{Taylor}, J.~H. 1974, AAPS, 15, 367

\bibitem[{{Tegmark} \& {Zaldarriaga}(2009)}]{tz09}
{Tegmark}, M., \& {Zaldarriaga}, M. 2009, \prd, 79, 083530

\bibitem[{{Tendulkar} {et~al.}(2017){Tendulkar}, {Bassa}, {Cordes}, {Bower},
  {Law}, {Chatterjee}, {Adams}, {Bogdanov}, {Burke-Spolaor}, {Butler},
  {Demorest}, {Hessels}, {Kaspi}, {Lazio}, {Maddox}, {Marcote}, {McLaughlin},
  {Paragi}, {Ransom}, {Scholz}, {Seymour}, {Spitler}, {van Langevelde}, \&
  {Wharton}}]{tbc+17}
{Tendulkar}, S.~P., {Bassa}, C.~G., {Cordes}, J.~M., {et~al.} 2017, \apjl, 834,
  L7

\bibitem[{{Thornton} {et~al.}(2013){Thornton}, {Stappers}, {Bailes},
  {Barsdell}, {Bates}, {Bhat}, {Burgay}, {Burke-Spolaor}, {Champion}, {Coster},
  {D'Amico}, {Jameson}, {Johnston}, {Keith}, {Kramer}, {Levin}, {Milia}, {Ng},
  {Possenti}, \& {van Straten}}]{tsb+13}
{Thornton}, D., {Stappers}, B., {Bailes}, M., {et~al.} 2013, Science, 341, 53

\bibitem[{{Trotter} {et~al.}(2017){Trotter}, {Reichart}, {Egger},
  {St{\'y}blov{\'a}}, {Paggen}, {Martin}, {Dutton}, {Reichart}, {Kumar},
  {Maples}, {Barlow}, {Berger}, {Foster}, {Frank}, {Ghigo}, {Haislip},
  {Heatherly}, {Kouprianov}, {LaCluyz{\'e}}, {Moffett}, {Moore}, {Stanley}, \&
  {White}}]{tre+17}
{Trotter}, A.~S., {Reichart}, D.~E., {Egger}, R.~E., {et~al.} 2017, \mnras,
  469, 1299

\bibitem[{{Vander Wiel} {et~al.}(2016){Vander Wiel}, {Burke-Spolaor},
  {Lawrence}, {Law}, \& {Bower}}]{vbl+16}
{Vander Wiel}, S., {Burke-Spolaor}, S., {Lawrence}, E., {Law}, C.~J., \&
  {Bower}, G.~C. 2016, ArXiv e-prints, arXiv:1612.00896

\bibitem[{{Yang} \& {Zhang}(2016)}]{yz16}
{Yang}, Y.-P., \& {Zhang}, B. 2016, \apjl, 830, L31

\bibitem[{{Yao} {et~al.}(2017){Yao}, {Manchester}, \& {Wang}}]{ymw17}
{Yao}, J.~M., {Manchester}, R.~N., \& {Wang}, N. 2017, \apj, 835, 29

\bibitem[{{Yoshiura} \& {Takahashi}(2018)}]{yt18}
{Yoshiura}, S., \& {Takahashi}, K. 2018, \mnras, 473, 1570

\bibitem[{{Zackay} \& {Ofek}(2017)}]{zo17}
{Zackay}, B., \& {Ofek}, E.~O. 2017, \apj, 835, 11

\end{thebibliography}

\end{document}